\documentclass[aps,prl,showpacs,twocolumn,amsmath,amssymb]{revtex4}
\usepackage{graphicx} 
\usepackage{dcolumn} 
\usepackage{bm} 

\begin{document}
\title{Spectral responses in granular compaction}
\author{Ling-Nan Zou}\email{zou@uchicago.edu}
\affiliation{The James Franck Institute and Department of Physics, The University of Chicago, Chicago, IL 60637}
\date{\today}

\begin{abstract}
The slow compaction of a gently tapped granular packing is reminiscent of the low-temperature dynamics of structural and spin glasses. Here, I probe the dynamical spectrum of granular compaction by measuring a complex (frequency-dependent) volumetric susceptibility $\tilde{\chi}_v$. While the packing density $\rho$ displays glass-like slow relaxations (aging) and history-dependence (memory) at low tapping amplitudes, the susceptibility $\tilde{\chi}_v$ displays very weak aging effects, and its spectrum shows no sign of a rapidly growing timescale. These features place $\tilde{\chi}_v$ in sharp contrast to its dielectric and magnetic counterparts in structural and spin glasses; instead, $\tilde\chi_v$ bears close similarities to the complex specific heat of spin glasses. This, I suggest, indicates the glass-like dynamics in granular compaction are governed by statistically rare relaxation processes that become increasingly separated in timescale from the typical relaxations of the system. Finally, I examine the effect of finite system size on the spectrum of compaction dynamics. Starting from the ansatz that low frequency processes correspond to large scale particle rearrangements, I suggest the observed finite size effects are consistent with the suppression of large-scale collective rearrangements in small systems.
\end{abstract}
\maketitle

\section{Randomly packed grains as a model glass}

Filling a container by pouring in spherical grains nearly always leads to a disordered packing. The ``ground state'' of the system, a face-centered cubic packing and the densest arrangement of spheres, is rarely achieved unless put in by hand. Shaking or tapping the container, causing the contents to settle and compact, works only so well: the packing remains disordered, and its packing density $\rho$ approaches but does not exceed the random close packing (RCP) density $\rho_\mathrm{rcp} \approx 0.64$ \cite{AsteWeaire2008, JaegerNagel1992, Scott1960}. Despite being a robust state that is readily reproducible, little is known about the origins of random closed packing.

\begin{figure}[tbp]
\center
\includegraphics[width= 0.45\textwidth]{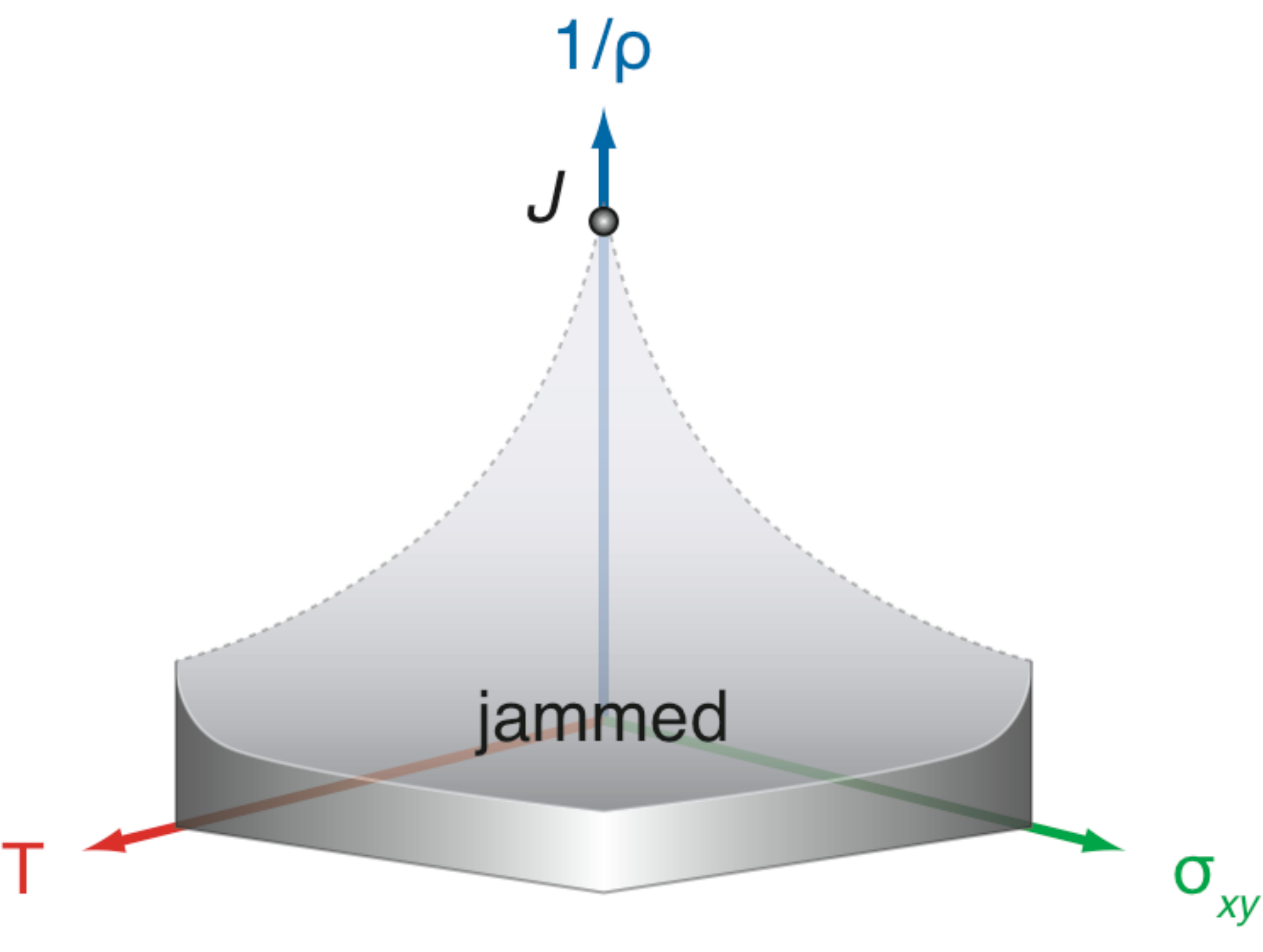}
\caption[The Jamming Phase Diagram]{(Color online) The Jamming Phase Diagram spanned by orthogonal axes inverse density $1/\rho$, temperature $T$, and shear stress $\sigma_{xy}$. The shaded volume is occupied by jammed or glassy states; the rest is filled by fluid states. Point $J$ marks the $T=0, \rho=\rho_\mathrm{rcp}$ jamming transition.}
\label{JammingPD}
\end{figure}

Surprisingly, the detailed arrangement of particles in RCP is remarkably similar to the molecular arrangements in a simple liquid \cite{Bernal1959, Bernal1960, Bernal1962, Scott1962, Finney1970a, Finney1970b}; but unlike a liquid, RCP is mechanically rigid and resists shear --- it is jammed \cite{Maxwell1864}. The same combination of rigidity and disorder also characterizes structural glasses, such as window glass and hard candies. Like RCP, a glass is an elastic solid whose molecular arrangement is nearly indistinguishable from that of a liquid. Indeed, the structure of simple metallic glasses are well-modelled by RCP \cite{Finney1977, Cahn1980, Zallen1998}. Drawing on these similarities, Liu and Nagel proposed that jammed packing of particles and glass-formation in simple liquids may be placed into a unified Jamming Phase Diagram spanned by inverse density $1/\rho$, temperature $T$, and shear stress \ (Fig.~\ref{JammingPD}) \cite{LiuNagel1998}. The region of high density, low temperature, and small shear is occupied by jammed or glassy states; the rest are filled by fluid states. Jamming is the consolidation of loose grains into a rigid packing when $\rho$ is increased at $T=0$, whereas the glass transition is the freezing of a liquid into a glassy solid at a finite $T$. The principal aim of this work is to explore the relationship between jammed grains and glasses by adopting methods and principles from glass experiments to study granular materials, in an effort to identify shared features, as well as differences and new effects.

\begin{figure}[tbp]
\center
\includegraphics[width= 0.45\textwidth]{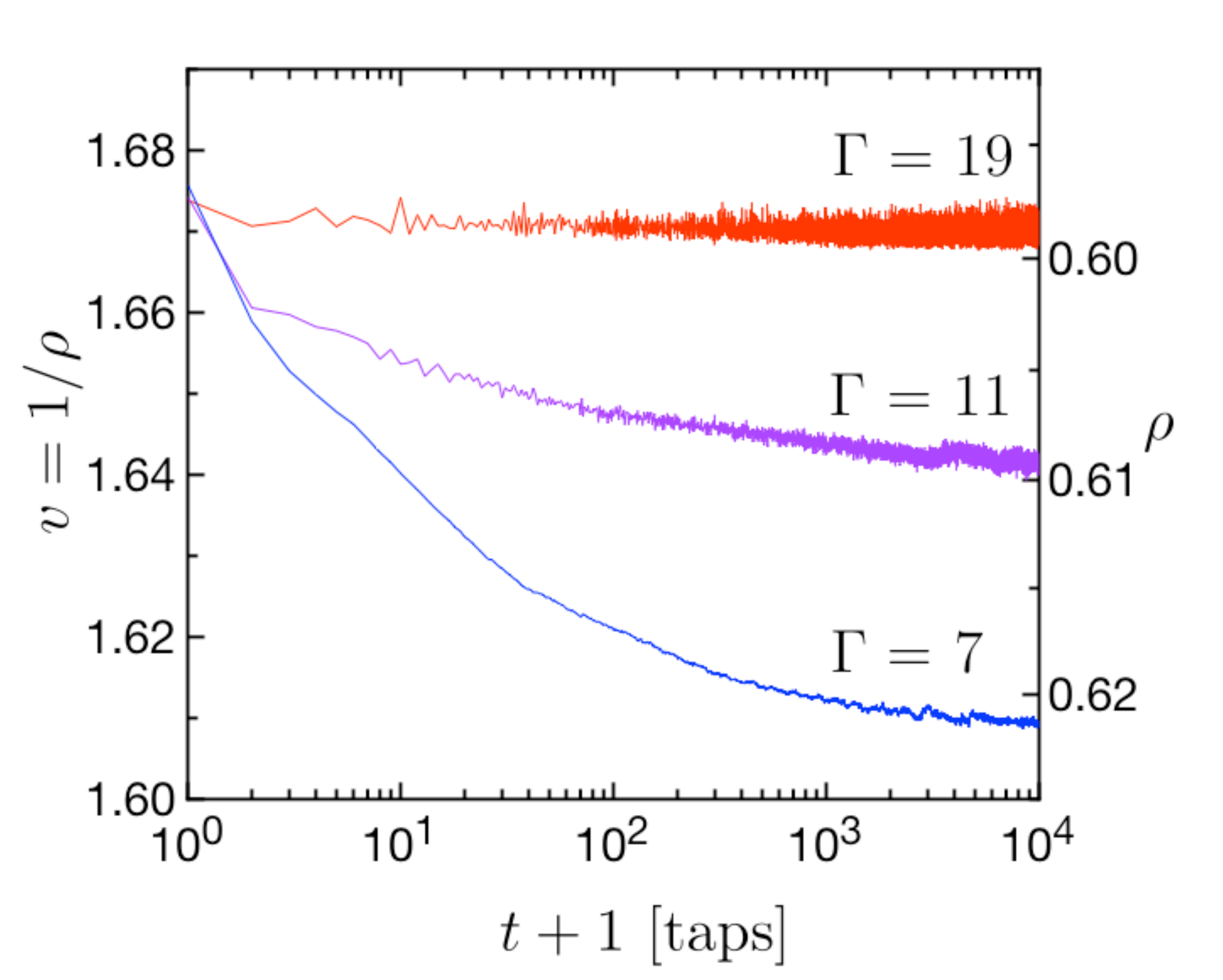}
\caption[Granular compaction under steady tapping]{(Color online) Granular compaction under steady tapping; $v\equiv 1/\rho$ is the specific packing volume, $t$ is the number of taps applied, and $\Gamma$ is the tapping amplitude (see Chapter 2.1). Strongly tapped ($\Gamma$ large), the packing quickly reaches a steady state; weakly tapped ($\Gamma$ small), the packing volume compacts slowly, and the steady state is not accessible within the experimental timescale.}
\label{CompactionDynamics}
\end{figure}

It is attractive to think random closed packing and glasses, unsolved problems on their own, may in fact be closely connected. It also invites skepticism to think systems so different microscopically can exhibit close analogies. We are therefore encouraged to find that granular packing is not only glass-like in structure, it can also exhibit slow relaxations reminiscent of the low-$T$ dynamics of structural glasses, and of other ``glassy'' systems such as spin glasses and kinetically constrained computer models \cite{EdigerAngellNagel1996, BinderYoung1986, RitortSollich2003}. For glass-forming liquids such as glycerol or viscous $\textrm{SiO}_2$, small reductions in $T$ can lead to large (and apparently diverging) increases in the relaxation time (and the fluid viscosity). Below the glass transition temperature $T_\mathrm{g}$, the characteristic relaxation time of the liquid becomes greater than the experimental timescale: the liquid falls out of equilibrium and becomes a glass. Because it is out of equilibrium, the properties of the glassy state are not stationary but depend on the experimental timescale (aging) and on the thermal history of the system (memory). For a packing of macroscopic particles with strongly dissipative interactions (friction), thermal energy is unimportant, and the dynamical evolution of the system is driven by external energy input (such as tapping or shaking). Strongly shaken, a granular packing quickly reaches a loosely-packed steady state. Weakly tapped, the packing compacts slowly in a logarithmic fashion with time, and the steady state may not be accessible on an experimental timescale (Fig.~\ref{CompactionDynamics}); furthermore, the detailed evolution depends on how the packing was prepared \cite{Knight&c1995, Nowak&c1997, PhillippeBideau2002, Nowak&c1998, Josserand&c2000, Richard&c2005}.

One may therefore take the random packing of grains --- despite its athermal, dissipative nature --- to be a model ``glass'', in terms both of structure and of dynamics. This raises the question whether there are shared physical principles connecting the various microscopically disparate systems commonly regarded as ``glassy''. Simply put: is a granular packing glassy in the same sense structural and/or spin glasses are glassy? If there is a deep connection, then granular packings may be a convenient experimental system for investigating the origins of glassy behavior. Here, I have developed a volumetric spectroscopy, modeled after the dielectric and magnetic spectroscopies of structural and spin glasses, to directly probe the spectrum of relaxation processes in granular compaction, with goal of ultimately comparing it, in detail, with the spectrum of glasses. Because granular packing is an athermal system without a well-defined statistical mechanics, analogies with the behavior of glasses at finite $T$ cannot be justified \textit{a priori}. I will adopt a purely operational approach: analogies will be proposed na\"ively, and they will be justified solely on the light they may shed on the experimental data. 

This paper is organized as follows:

Section II describes the volumetric spectroscopy and gives an operational definition for the ``susceptibility'' $\tilde{\chi}_v$. I will then describe how the spectroscopy is performed, and $\tilde{\chi}_v$ measured, in a granular compaction experiment.

Section III describes the behavior of $\tilde{\chi}_v$ as a function of the taping amplitude (which is the principal control parameter for granular compaction) and the frequency (the dynamical spectrum). Despite some similarities to the dielectric/magnetic susceptibilities of structural and spin glasses, $\tilde{\chi}_v$ is distinctive in exhibiting very weak aging. More surprisingly, there is no indication in the spectrum of $\tilde{\chi}_v$ of a diverging relaxation time as the tapping amplitude is turned down. Comparing the compaction process with the dynamical spectrum, one finds that granular compaction is characterized by a broad separation of timescales.

Section IV suggest that the apparently distinctive behavior of $\tilde{\chi}_v$ may be generic to a complex configurational specific heat. In analogy to the well-known relationship between specific heat and energy fluctuations, the relationship between $\tilde\chi_v$ and specific volume fluctuation is found to be consistent with an effective Fluctuation-Dissipation Theorem. Furthermore, computer simulations of two standard spin glass models found the complex specific heat in these models appear to mirror the behavior of $\tilde{\chi}_v$. If  $\tilde{\chi}_v$ can indeed be taken as a ``specific heat'', it suggest the slow dynamics of granular compaction is governed by statistically rare relaxation processes which become increasing decoupled from the typical dynamics of the system.

Section V describes how the system size affects the dynamics of granular compaction. For small systems, $\tilde{\chi}_v$ is suppressed within an intermediate frequency band and within an intermediate window of tapping amplitude; this is consistent with the suppression of large-scale correlated motions in small systems. But for sufficiently weak driving, the low-frequency responses of small systems are enhanced; I suggest this is the consequence of particle/wall interactions at the boundary.

To conclude, in Section VI, I will summarize the most salient results and briefly discuss their implications for understanding ``glassy'' dynamics. I will close by suggesting ways that this work can be extended to address several important open questions.

\section{The volumetric spectroscopy of granular compaction}

\begin{figure*}[tbp]
\center
\includegraphics[width= 0.945\textwidth]{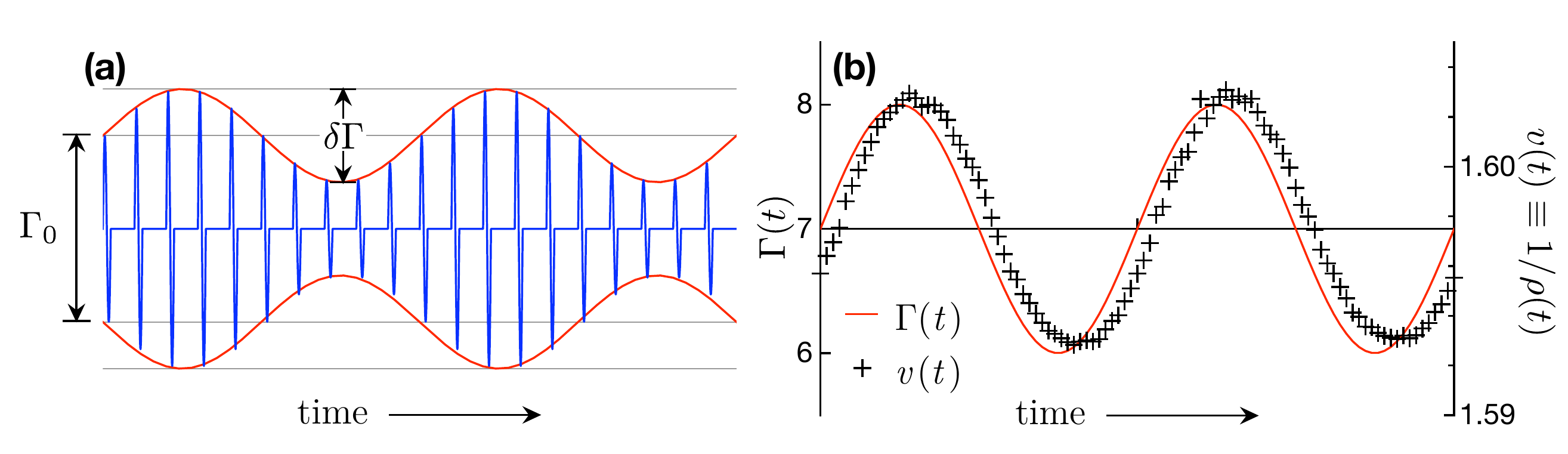}
\caption[Amplitude-modulated tapping]{(Color online) (a) Amplitude modulation of discrete taps. (b) Input: $\Gamma(t)$, amplitude-modulated tapping as a function of the tap number $t$ (solid line). Experimentally measured response: frequency-locked modulation in the specific packing volume $v(t)$ (filled circles); here the time series has been divided into two time-blocks with lengths of $1/f=50$ taps. Note that $\Gamma(t)$ and $v(t)$ are slightly out of phase.}
\label{AM}
\end{figure*}  

The dynamics of structural and/or spin glasses is commonly studied via an appropriate spectroscopy. Here, temperature $T$ is usually the principal parameter that controls whether the system is in the glassy state. Having set the temperature, the dynamics of system is characterized by some susceptibility, which describes how the system responds to small perturbations. For example, the dielectric susceptibility $\epsilon$, often used to characterize glass-forming liquids, describes the response of the net polarization $\mathbf{P}$ when a small electric field $\mathbf{E}$ is applied: $\epsilon = \partial \mathbf{P}/\partial \mathbf{E}$. If the perturbation can be applied in a frequency-controlled way, e.g. with an AC electric field, a dynamical spectrum of response versus frequency can be obtained. Such a spectrum will encode the relative distribution of relaxation times in the system.

In a similar spirit, I will characterize granular compaction using a complex (frequency-dependent) volumetric ``susceptibility'' $\tilde{\chi}_{v}$, which describes the response of the specific packing volume $v\equiv 1/\rho$ to small changes in the external driving (tapping), and extract the dynamical spectrum of granular compaction.  This should allow us to make a detailed comparison between the slow compaction of a granular packing and the low-temperature dynamics of structural and spin glasses.

\subsection{An operational definition of $\tilde\chi_{v}$}

In a typical compaction experiment, a granular packing (held inside a container) is compacted with a sequence of identical taps, separated sufficiently apart so that the packing comes to a complete rest before being tapped again. Each tap is one single period of sinusoidal (30 Hz, in my case) vertical vibration, and the tapping amplitude $\Gamma$ is given by the peak-to-peak acceleration $a_\mathrm{pp}$ of the tap, normalized by $g=9.8 \ \textrm{m}/\textrm{s}^2$: $\Gamma\equiv a_\mathrm{pp}/g$. In order to extract a susceptibility, the packing must be perturbed at the same time it is being tapped; however, experimentally there is no readily accessible secondary field which can couple to the system.

The solution here is to use tapping both as the principal control parameter and as the perturbing ``field''. Instead of a sequence of taps with identical amplitudes, the tapping amplitude is modulated about a mean level [Fig~\ref{AM}(a)],
\begin{equation}
\Gamma(t) = \Gamma_0 + \delta\Gamma\sin (2\pi f t).
\end{equation}
The mean amplitude $\Gamma_0$ serves as the principal control parameter, while the modulation amplitude $\delta\Gamma$ serves as the perturbing ``field''; $f$ is the modulation frequency in inverse taps, and $t$ is time measured in the number of taps applied. In response to the modulated tapping, the temporal evolution of the specific packing volume $v\equiv 1/\rho$ can be described as the sum of a slowly varying component $v_0$, and a frequency-locked component which is phase-shifted with respect to the input [Fig~\ref{AM}(b)],
\begin{equation}
v(t) \approx v_0 + \delta v\sin (2\pi f t - \varphi).
\end{equation}
Here $\delta v$ is the response magnitude, and $\varphi$ is the phase shift between the modulated tapping and the volumetric response. One can then define the real and imaginary (in- and out-of-phase) parts of a complex susceptibility $\tilde{\chi}_v = (\chi_v', \chi_v'')$ as
\begin{eqnarray}
\chi_v' &=& \left(\frac{\delta v}{\delta\Gamma}\right) \cos \varphi, \label{chi_def1} \\
\nonumber \\
\chi_v'' &=& \left(\frac{\delta v}{\delta\Gamma}\right) \sin \varphi. \label{chi_def2}
\end{eqnarray}
This susceptibility $\tilde\chi_{v}$ describes how the specific volume $v$ responds to a small change in the tapping amplitude $\delta\Gamma$. For sufficiently weak perturbations, the response should be linear, $\delta v \propto \delta\Gamma$, and $\tilde\chi_v$ should be independent of the modulation aamplitude.

In practice, the phase shift $\varphi$ and the response amplitude $\delta v$ are not separately extracted from the $v(t)$ data. Using the orthonormality of trigonometric functions, $\tilde{\chi}_{v}$ can be directly extracted by projecting $v(t)$ onto its in- and out-of-phase components (with respect to the tapping modulation). Once a $v(t)$ time series has been collected, it is divided into time-blocks of length $1/f$, synchronized to the tapping modulation. Within each time-block, $v(t)$ is discretely transformed to extract $\tilde\chi_{v}$ directly:
\begin{equation}
\chi_{v}'(t) = \frac{2f}{\delta\Gamma} \sum_{\tau=0}^{1/f - 1} v(t+\tau) \sin(2\pi f \tau),
\label{in_phase}
\end{equation}
\begin{equation}
\chi_{v}''(t) =\frac{2f}{\delta\Gamma}  \sum_{\tau=0}^{1/f - 1} v(t+\tau) \sin(2\pi f \tau - \pi/2).
\label{out_phase}
\end{equation}
Formally, Eqs.~\ref{in_phase}, \ref{out_phase} are exact projections only if the background $v_{0}$ of $v(t)$ is stationary. However, since $v_{0}$ is almost always slowly-varying on the timescale of $1/f$, Eqs.~\ref{in_phase}, \ref{out_phase} are excellent approximations.

\subsection{Experimental design}

\begin{figure}[tbp]
\center
\includegraphics[width= 0.45\textwidth]{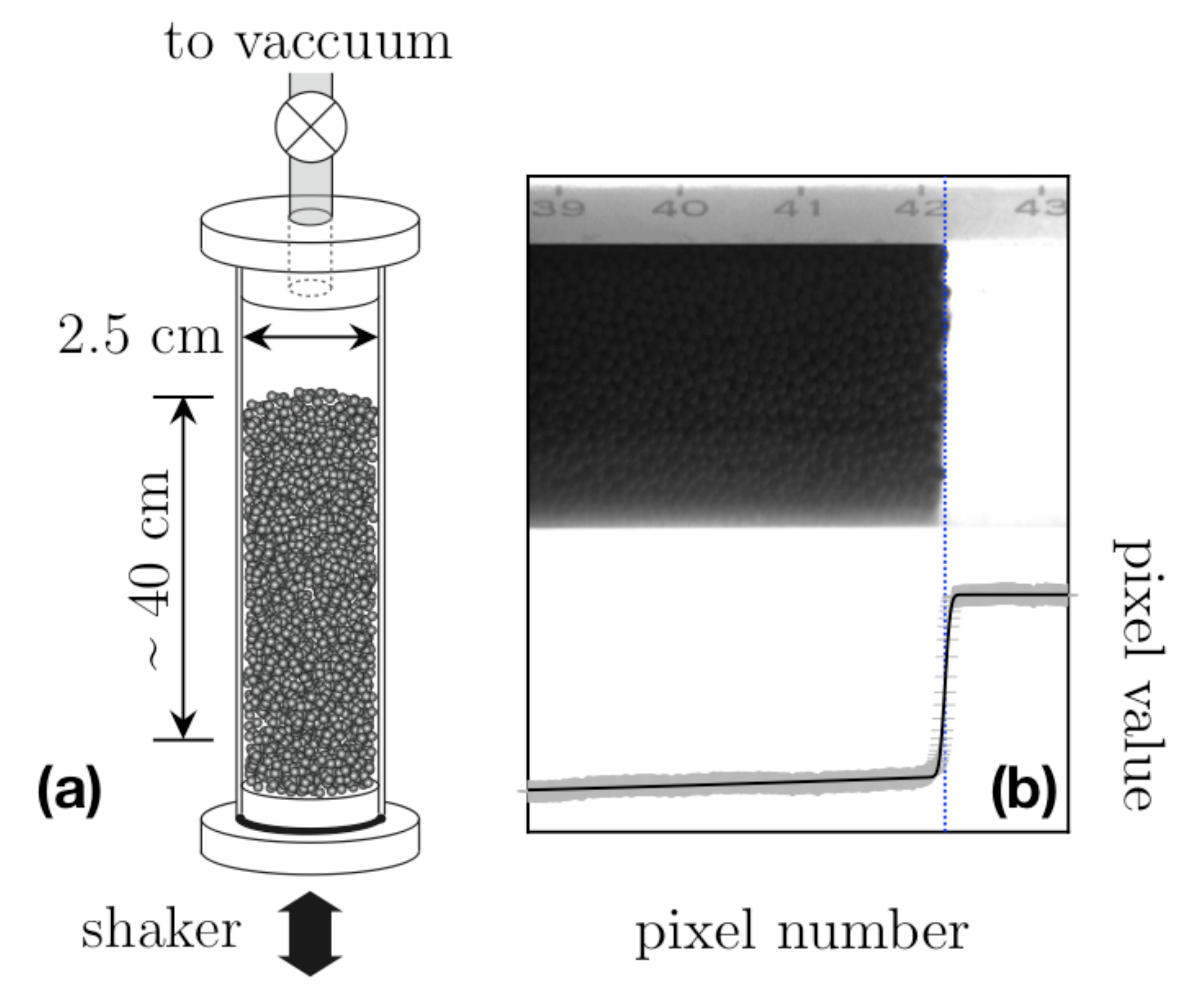}
\caption[The experimental setup]{(a) The experimental setup used to measure $\tilde\chi_v$ (schematic not drawn to scale). (b) A photograph of the packing; its image profile is fitted to extract the packing height, and hence the specific packing volume $v$.}
\label{Experiment}
\end{figure}

The experimental design here is partly modeled after that of ref~\cite{Knight&c1995}. I use a cylindrical packing whose height is much larger than its diameter in an effort suppress large scale convections during the compaction process. The cell is a smooth-walled cylindrical tube made of polycarbonate, 50 cm tall, with an inner diameter $D=2.5$ cm. For experiments, the cell is filled with particles to a height of $\approx 40$ cm [Fig.~\ref{Experiment}(a)]. To minimize static build-up, the inside cell wall is coated with a hard conductive polymer coating (Agfa Orgacon S103, ITO America Corp., Tempe, AZ). The cell is mounted on a computer-controlled electromechanical shaker, which is carefully leveled and set on a vibration-damping platform. To minimize unwanted lateral motions during tapping, the experimental cell is braced to a stiff external framework with heavy-duty rubber bands. 

The particles used are spherical zirconium oxide/silica ceramic (material density = $3.88\ \textrm{g/cm}^{3}$) particles (Glenn Mills Inc., Clifton, NJ). Compared to more commonly used glass beads, these are mechanically much tougher. They also have a much smoother surface, which minimizes abrasion on the cell and on the particles themselves. These features help to ensure good reproducibility over the course of very long experiments and millions of taps. Particles of three different diameters were used: $d=0.5$ mm, 1.1 mm, and 1.8 mm, all with $\sim 10\%$ polydispersity. Because air effects can significantly alter the compaction process \cite{PakVanDoornBehringer1995}, particularly when particles are small, once loaded the cell is sealed and evacuated to a pressure $<5 \ \textrm{Pa}$.

To measure the specific packing volume, a photograph of the packing is taken after each tap using a computer-controlled digital camera running Astro IIDC (Outcast Software, Calgary, Alberta). The cell is backlit with a uniform, bright light source. Where the cell is filled with particles (which are opaque), the image is completely dark. By fitting the intensity profile of the image to an error function, the location of the top surface of the packing can be found [Fig.~\ref{Experiment}(b)]. This is then compared with known reference marks on the cell to extract the packing height, and hence, the packing volume. The absolute accuracy in the determination of $v$ is limited by the unevenness is the top surface and by the accuracy with which reference marks can be placed. These are compensated by the large fill height, which keeps the absolute uncertainty in $v$ to $\sim 5\times10^{-3}$. The relative accuracy, which is what matters most here (since the signal I seek consists of small changes about a slowly-varying background), is somewhat better ($\sim 10^{-4}$) due to the fact that image profiles can be tracked with sub-pixel resolution. Compared to methods that measure the packing density locally, such as capacitance \cite{Knight&c1995} or $\gamma$-ray absorption \cite{PhillippeBideau2002}, the imaging technique measures the bulk average and therefore displays much less fluctuation from tap to tap. This is necessary in order to obtain a clean measurement of $\tilde\chi_v$.

To ensure a fresh start at for each new experiment, the packing is ``reset'' by tapping at $\Gamma = 21$ for at least $10^{3}$ taps. The result is a reproducible loose packing that has achieved a steady-state, where all memory of its prior tapping history appears to be lost. Unless noted otherwise, this is the loose-packed initial state I will use for each new experimental run.

\section{Characterizing the volumetric response}

The volumetric susceptibility $\tilde{\chi}_{v}$ depends on both $\Gamma_0$, which is the principal control parameter, and on the modulation frequency $f$. It should not depend on $\delta\Gamma$ if one is in the linear response regime, i.e. $\delta v \propto \delta\Gamma$. At low $\Gamma_0$, where granular compaction exhibits glassy behavior, $\tilde{\chi}_{v}$ should in principle depend on the experimental timescale (aging), and the previous history of the packing (memory). Indeed this is what we expect based on the behavior of dielectric and magnetic susceptibilities in structural and spin glasses. Unless explicitly noted, all the results described in this Section were obtained using $d=0.5$ mm particles, giving an effective system size of $D/d = 50$.

\subsection{Dependence of $\tilde{\chi}_{v}$ on $\Gamma_0$ and $\delta\Gamma$}

\begin{figure}[tbp]
\center
\includegraphics[width= 0.45\textwidth]{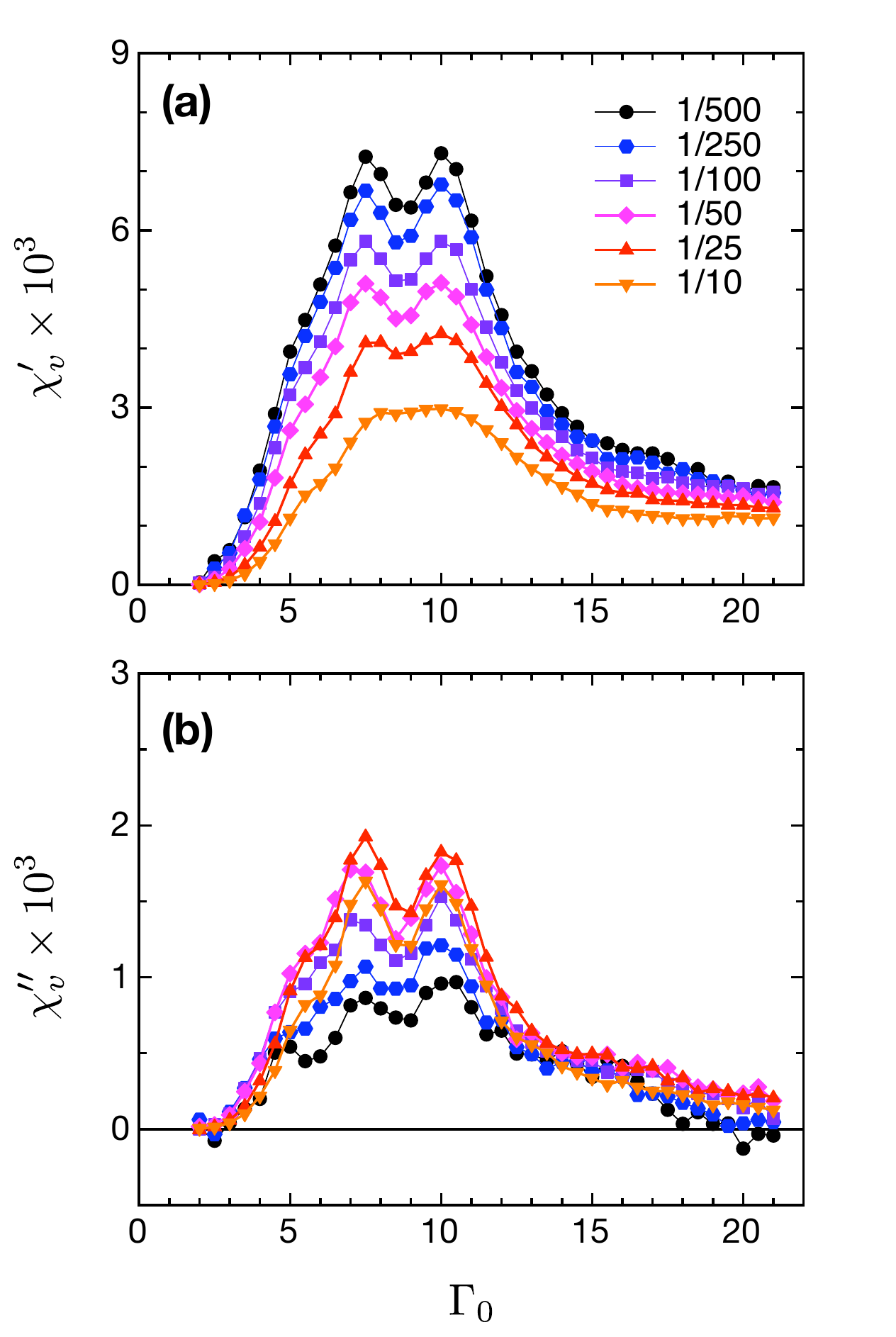}
\caption[The dependence of $\tilde\chi_{v}$ on $\Gamma_{0}$]{(Color online) The (a) real and (b) imaginary parts of the volumetric susceptibility $\tilde\chi_v$ as a function of $\Gamma_0$, measured at different modulation frequencies $f$.}
\label{chi(Gamma)}
\end{figure} 

\begin{figure}[tbp]
\center
\includegraphics[width= 0.45\textwidth]{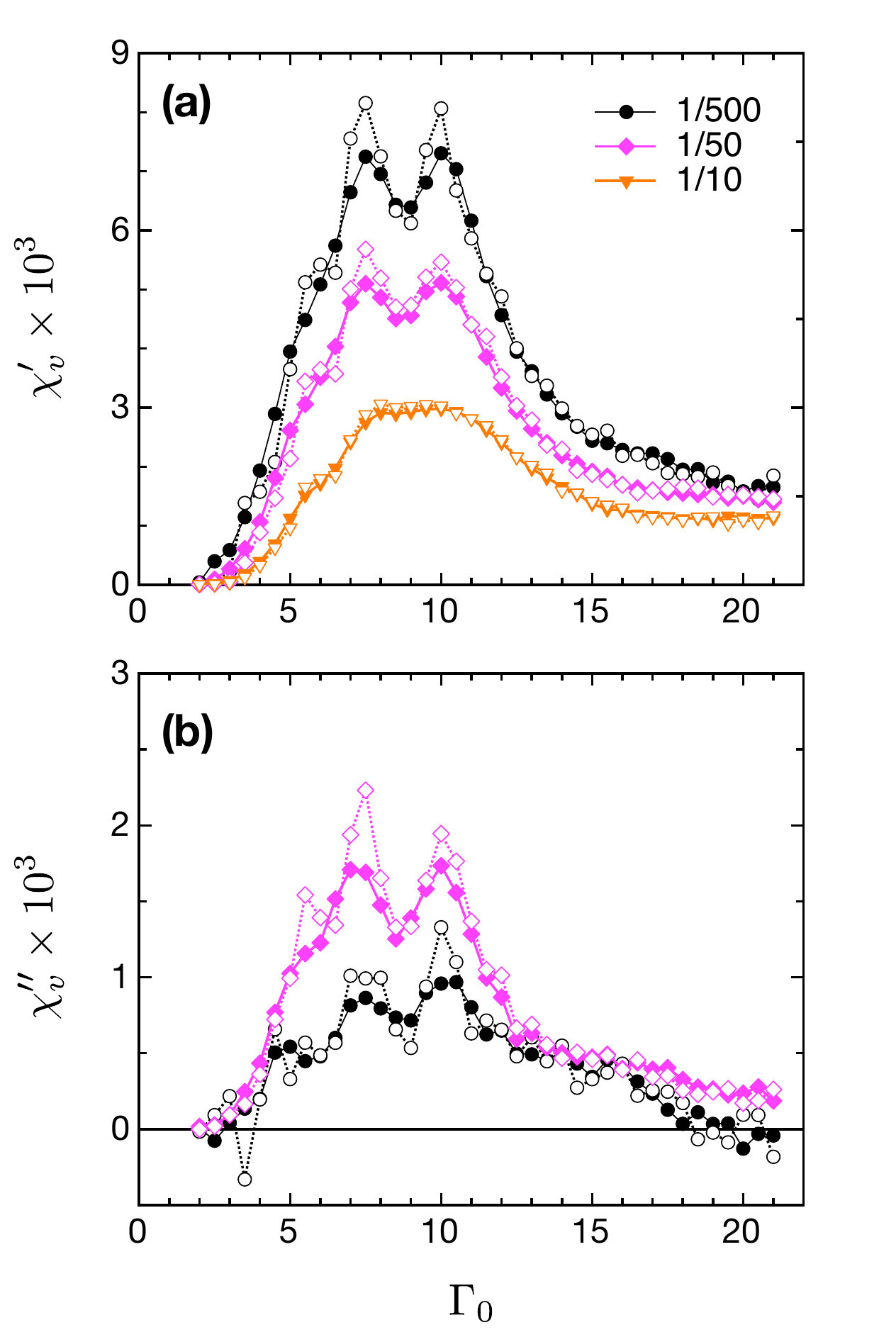}
\caption[Testing the linear response of $\tilde\chi_{v}$ with $\delta\Gamma$]{(Color online) Testing for linear response: (a) $\chi_v'$ and (b) $\chi_v''$ measured using $\delta\Gamma=1$ (filled symbols) and $\delta\Gamma=0.5$ (open symbols) at several different frequencies (as indicated in figure legend). The similarity in the resulting curves shows the experiment is in the linear response regime.}
\label{chi(Gamma)LR}
\end{figure} 

To measure $\tilde{\chi}_v$ as a function of $\Gamma_0$ (at a fixed frequency $f$), the mean tapping amplitude is ramped from $\Gamma_0=21$ to $\Gamma_0=2$ and cycled back in steps of $\Delta\Gamma_0 = 0.5$ and a dwell time per step of $\tau_\mathrm{dw} = 500$ taps. Fig.~\ref{chi(Gamma)} plots $\tilde{\chi}_v(\Gamma_0)$ measured at different frequencies, where the modulation amplitude is set to $\delta\Gamma = 1$. Because little hysteresis is apparent during the cycling, I have averaged together the values of $\tilde{\chi}_{v}$ obtained during the down-ramp with that obtained during the up-ramp. As a function of $\Gamma_0$, both $\chi_v'$ and $\chi_v''$ vary non-monotonically with $\Gamma_0$, exhibiting a broad maximum at an intermediate range of $\Gamma_0$ for all the frequencies measured. This shows intermediate $\Gamma_{0}$ is where compaction is most sensitive to variations in the tapping amplitude. The broad maximum is structured, with sub-peaks at $\Gamma_0=10$ and $\Gamma_0=7.5$, and a ``knee'' at $\Gamma_0=5$. These sub-features are found in both $\chi_v'$ and $\chi_v''$, and are located at the same values of $\Gamma_0$. At lower frequencies, one obtains a stronger the real response, and the fine features in $\chi_{v}(\Gamma_{0})$ stand out more pronouncedly. On the other hand, the imaginary response at low frequencies is quite weak.

As the mean tapping amplitude $\Gamma_0$ and the temperature $T$ are, respectively, the principal control parameters for granular compaction and for glass formation, it is reasonable to compare $\tilde{\chi}_{v}(\Gamma_0)$ with the $T$-dependent susceptibilities of glasses. Qualitatively, $\tilde{\chi}_{v}(\Gamma_0)$ has broad similarities to the magnetic susceptibility $\tilde{\chi}_{M}(T)$ of spin glasses, such as CuMn, as well as the dielectric susceptibility $\tilde\epsilon(T)$ of structural glasses such as glycerol \cite{Menon&c1992, MuldervanDuyneveldtMydosh1981, MuldervanDuyneveldtMydosh1982}. However, for most structural and spin glasses, the peak in $\tilde\epsilon(T)$ or $\tilde{\chi}_{M}(T)$ has a simple structure, whereas $\tilde{\chi}_{v}(\Gamma_0)$ exhibits multiple sub-peaks. For glasses, the peak in $\tilde\epsilon(T)$ or $\tilde{\chi}_{M}(T)$ shifts to lower $T$ for lower frequencies. In granular compaction, the fine features $\tilde\chi_{v}(\Gamma_{0})$ remains at the same values of $\Gamma_{0}$.

Fig.~\ref{chi(Gamma)LR} compares $\tilde{\chi}_{v}(\Gamma_0)$ measured using $\delta\Gamma=1$ with that obtained using a smaller modulation amplitude $\delta\Gamma=0.5$; the results are practically identical. This indicates the experiments are largely in the linear response regime within the experimental range of $f$ and $\Gamma_0$. On closer examination, the sub-features in $\tilde{\chi}_{v}(\Gamma_0)$ are more pronounced at the smaller $\delta\Gamma$. This suggests the sub-peaks of $\tilde{\chi}_{v}$ mark values of $\Gamma_{0}$ where the packing is particularly sensitive, and where the response is most nonlinear. This is reminiscent of the extra sensitivity one expects of systems near a critical point. A similar effect can be found in spin glasses, where low-magnetic field (weaker perturbation) measurements yield a taller, sharper peak in the magnetic susceptibility. But where in the spin glass, the peak of $\tilde\chi_M$ shifts to higher $T$ when measured using smaller magnetic fields \cite{ZhouBakk1994, AnderssonJonssonMattsson1996}, in granular compaction the sub-peaks in $\tilde\chi_v$ do not shift with $\delta \Gamma$.

For structural or spin glasses, where the dielectric or magnetic susceptibility peaks is often taken to mark the glass transition temperature $T_{g}$. Here, we can ask whether the features in $\tilde{\chi}_{v}$ correspond to any ``transitions'' in compaction dynamics with $\Gamma_0$.

\begin{figure}[tbp]
\center
\includegraphics[width= 0.45\textwidth]{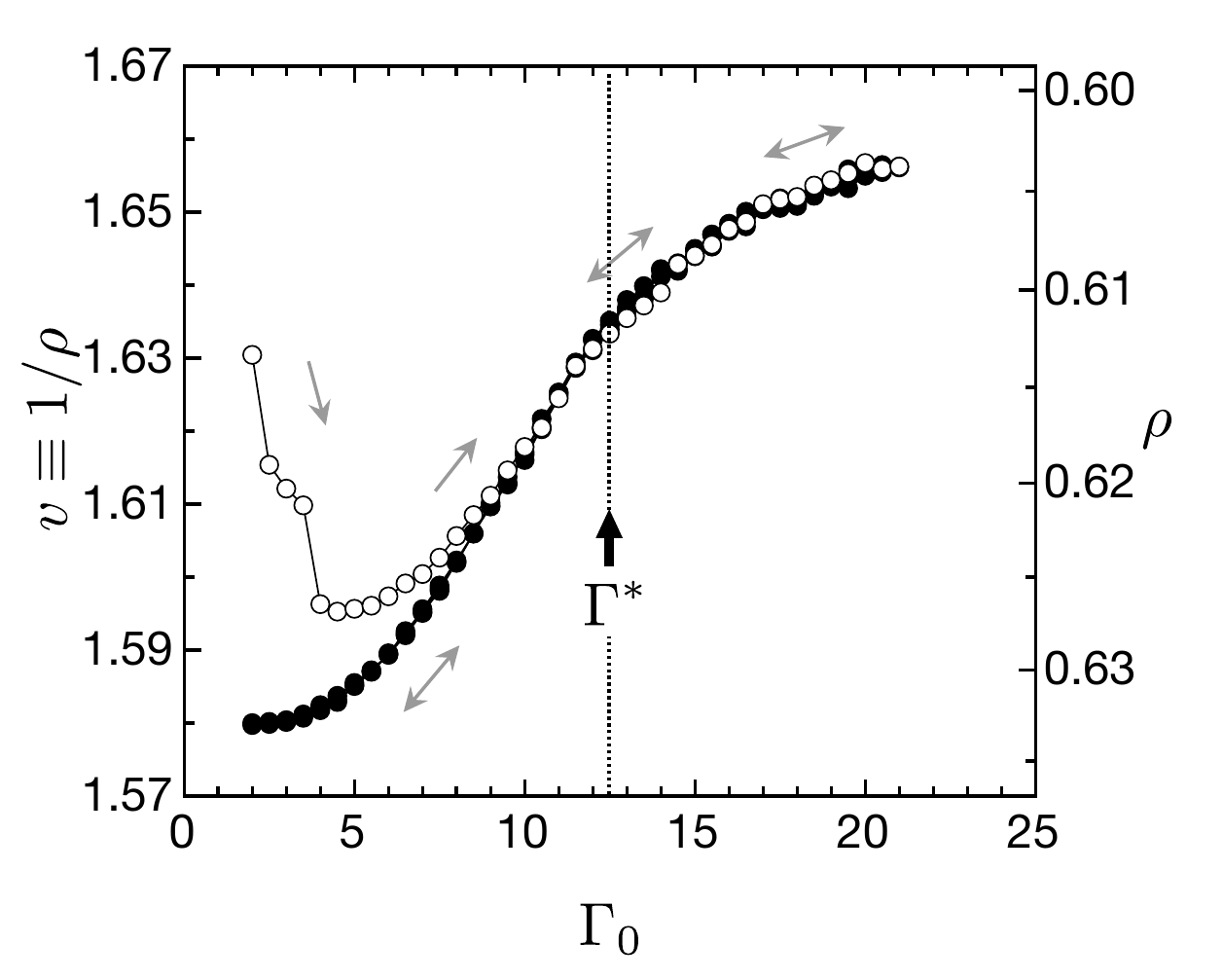}
\caption[The volumetric curve $v(\Gamma_{0})$]{The volumetric curve $v(\Gamma_0)$, showing the irreversible (open symbols) and reversible (filled symbols) branches. Arrows indicate the direction $\Gamma_0$ is ramped, and $\Gamma^*$ marks where the two branches appears to converge.}
\label{v(Gamma)Single}
\end{figure} 

A candidate location for such a transition is the branch point $\Gamma^{*}$, where the reversible and irreversible branches of the volumetric curve $v(\Gamma_0)$ converge \cite{Nowak&c1997}. As shown in Fig.~\ref{v(Gamma)Single}, starting from a loose-packed initial state and ramping up $\Gamma_{0}$ slowly from a small value, the packing initially compacts with increasing $\Gamma_{0}$, before it reverse course and begins to dilate with further increases of $\Gamma_{0}$; this trajectory is the ``irreversible branch''. If $\Gamma_{0}$ is then slowly ramped down from large amplitude to small, the packing compacts monotonically, becoming denser at lower $\Gamma_{0}$; subsequent cycling of $\Gamma_{0}$ between strong and weak tapping (at the same ramping rate) will remain on this so-called reversible branch. For $\Gamma_{0}>\Gamma^*$, the two branches converge, and the packing reaches a steady state with a unique volume on the experimental timescale ($\sim$ inverse ramping rate). For $\Gamma_{0}<\Gamma^*$, $v$ depends not only on $\Gamma_{0}$ but also on prior tapping history. In this sense, $\Gamma^*$ is like $T_\mathrm{g}$ in that for $T<T_{\mathrm{g}}$, the properties of a glass depends on its  thermal history.  As the convergence of the irreversible and reversible branches is gradual, $\Gamma^*$ is a not precise location but depends (weakly) on the rate at which $\Gamma_{0}$ is ramped \cite{Nowak&c1997}. This is reminiscent of the fuzziness of $T_\mathrm{g}$, which depends weakly on the rate at which the glass-forming system is cooled  down from high $T$.

The susceptibility $\tilde{\chi}_{v}(\Gamma_0)$ can be thought as the complex ``derivative'' of the reversible branch of the compaction curve $v(\Gamma_{0})$. Thus the broadly peaked form of $\tilde{\chi}_{v}(\Gamma_{0})$ is not a surprise given the form of $v(\Gamma_{0})$. The sub-peak in $\tilde{\chi}_{v}$ at $\Gamma_0=10$ is located near, but slightly below $\Gamma^{*}$. But for the equally prominent peak located at $\Gamma_0=7.5$, there is no outstanding feature in $v(\Gamma_{0})$ that appears to correspond to it.

\subsection{Aging and memory effects in $\tilde\chi_{v}$}

\begin{figure}[tbp]
\center
\includegraphics[width= 0.45\textwidth]{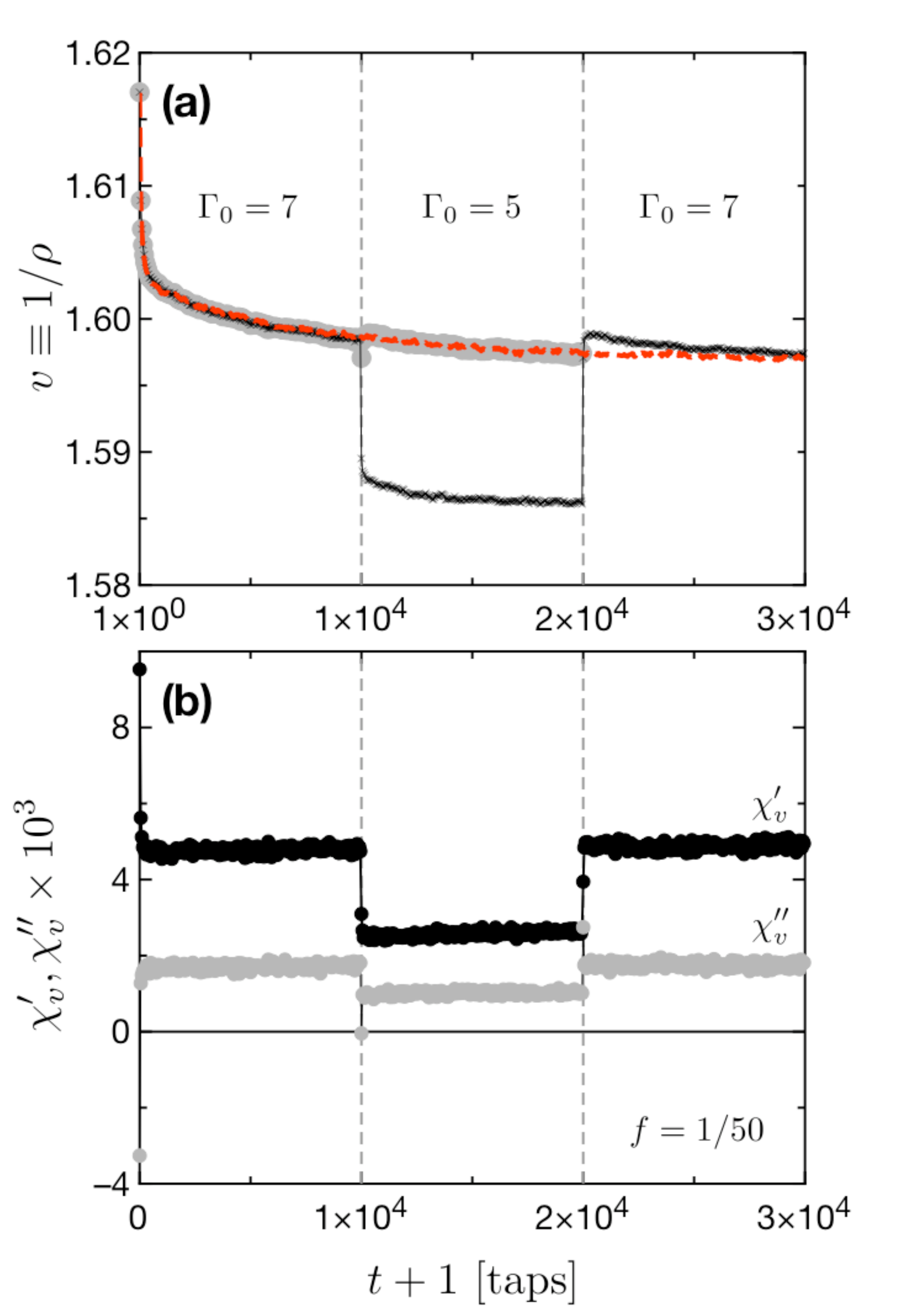}
\caption[Aging and memory in granular compaction]{(Color online) Granular compaction subjected to an step-wise shift protocol $\Gamma_{0} = 7\rightarrow 5\rightarrow 7$, starting from a loose-packed initial state. (a) Specific volume $v(t)$ (crosses), showing strong aging signatures both initially, and at the negative shift to $\Gamma_{0}=5$. When the two $\Gamma_{0}=7$ portions of $v(t)$ stitched together (grey-filled circles), the results is the same as an uninterrupted $\Gamma_{0}=7$ $v(t)$ reference curve (bold broken line),  aside from a brief transient. (b) The volumetric susceptibility $\tilde\chi_{v}$ measured in the same experiment. The aging signature is very weak, and the memory effect is trivial.}
\label{AgingMemory}
\end{figure} 

\begin{figure}[tbp]
\center
\includegraphics[width= 0.45\textwidth]{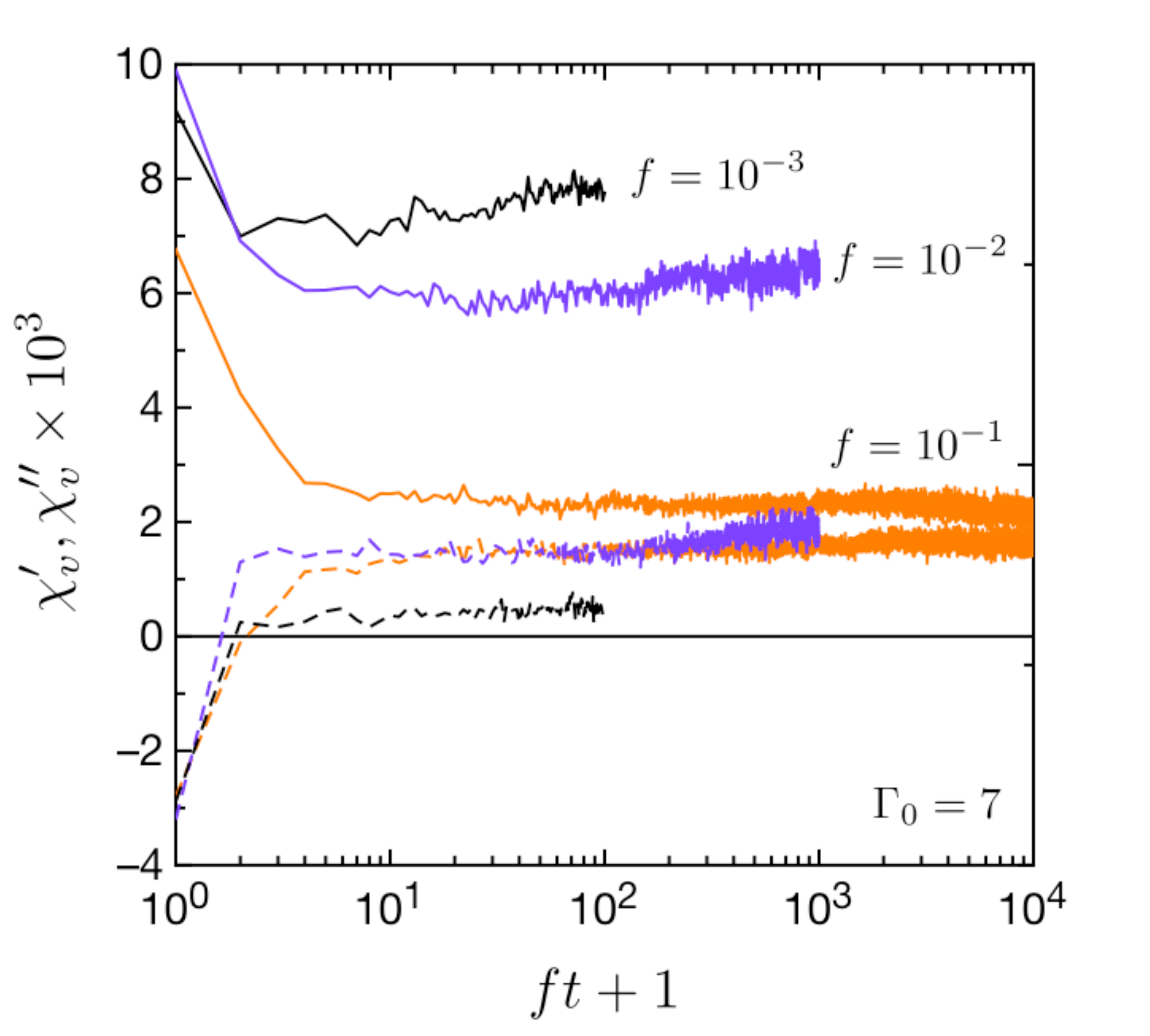}
\caption[Weak aging effects in $\tilde\chi_{v}$]{(Color online) The relaxation of $\chi_{v}'$ (solid lines), $\chi_{v}''$ (broken lines)  at $\Gamma_{0}=7$ for $f=10^{-1}$ (orange), $f = 100^{-1}$ (purple) , and $f=1000^{-1}$ taps$^{-1}$ (black). $\tilde\chi_{v}$ is mostly relaxed after a timescale on the order of $1/f$. }
\label{chiDecay}
\end{figure} 

Fig.~\ref{AgingMemory}(a) shows the behavior of granular compaction subjected to a step-wise shift protocol in $\Gamma_0$, starting from a loosely-packed initial state. With the initial onset of tapping at $\Gamma_0=7$, the specific volume $v$ exhibits a strong aging signature and slowly compacts in a quasi-logarithmic fashion with $t$. After $10^4$ taps, $\Gamma_0$ is turned down to $\Gamma_0=5$, and the system responds with renewed bout of aging and relaxes to a smaller volume. After $10^4$ taps at $\Gamma_0=5$, the system is returned to $\Gamma_0=7$, whereupon the packing dilates to a volume close to what it had just before the excursion to $\Gamma_0=5$. When the two $\Gamma_0=7$ portions of the data are stitched together, the result is identical (aside from a transient) to an uninterrupted $\Gamma_0=7$ aging curve. 

In structural and spin glasses, the experiment described above is carried out via step-wise shifts in the temperature $T$, and the aging and memory effects are observed in the dielectric or spin susceptibility \cite{LehenyNagel1998, Vincent&c1995}. It is a classic demonstration of both the aging and memory effects in a glassy system. For granular compaction, in the same experiment where aging and memory effects are so clearly displayed in the specific volume $v$, the volumetric susceptibility $\tilde{\chi}_v$ exhibits no discernible aging signature [Fig.~\ref{AgingMemory}(b)]. $\tilde{\chi}_{v}$ relaxes within a few modulation periods $1/f$ to apparently static values in response to the shifts in $\Gamma_0$. In this respect, $\tilde{\chi}_{v}$ behaves rather unlike dielectric and magnetic susceptibilities of structural and spin glasses with their strong aging effects. A closer examination shows that aging effects in $\tilde\chi_v$ are not entirely absent, but that they are very weak. Fig.~\ref{chiDecay} plots the relaxation of $\tilde{\chi}_v$ at different frequencies against the reduced timescale $ft$, starting from loosely packed initial states; here $\Gamma_0=7$, and $f=10^{-1}, 10^{-2}$, and $10^{-3} \ \textrm{taps}^{-1}$. The susceptibility appears to be largely relaxed after a timescale on the order of $1/f$. 

\subsection{The spectrum of $\tilde{\chi}_{v}$}

The weak aging effects in $\tilde{\chi}_v$ can be employed to our advantage in the measurement of its spectrum. Even if the system is not in steady state, the measured susceptibility, due to its weak aging effects, should be close to the ``true'' steady-state value. Thus the spectrum of $\tilde\chi_v$ can be measured down to fairly small $\Gamma_0$, where the steady state is experimentally inaccessible, with the confidence that the resulting spectrum will be close to the steady-state one. Here, starting from a loosely packed initial state, $\Gamma$ is slowly ramped stepwise (step size $\Delta\Gamma=0.5$, dwell time/step $\tau_\mathrm{dw} = 500$ taps) to the working $\Gamma_0$. The packing is then tapped at the working $\Gamma_0$ (without modulation) for an extended period until either steady state is achieved, or until the volumetric evolution is slow on the order of the lowest frequency to be probed. Only then is the amplitude modulation ($\delta\Gamma = 1$) turned on, and the frequency swept from high frequencies to low.

\begin{figure}[tbp]
\center
\includegraphics[width= 0.45\textwidth]{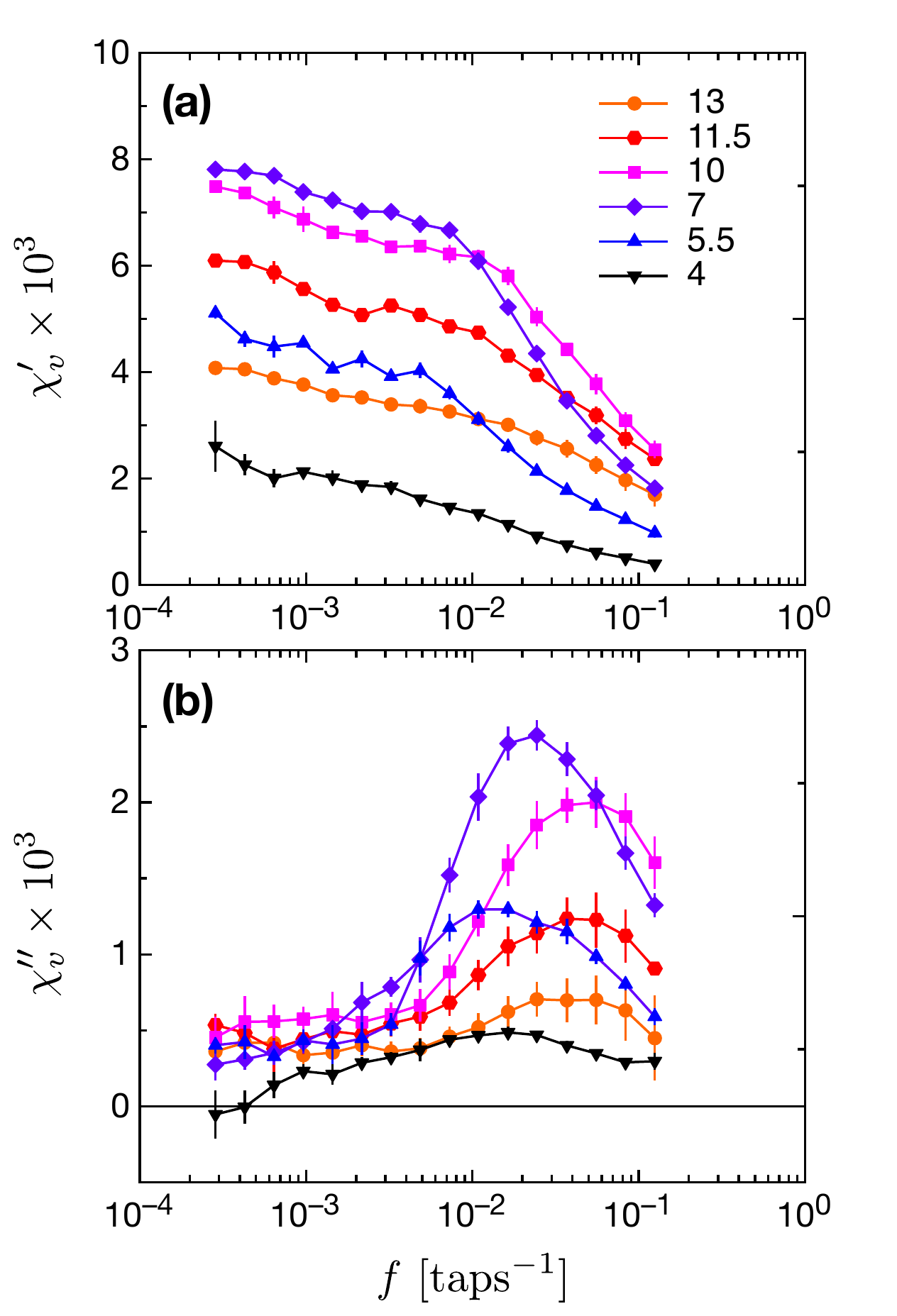}
\caption[The frequency spectrum of $\tilde\chi_{v}$]{(Color online) The (a) real and (b) imaginary parts of the $\tilde\chi_{v}$ spectrum, obtained at different values of $\Gamma_{0}$(as indicated in figure legend).}
\label{Spectra}
\end{figure} 

\begin{figure}[tbp]
\center
\includegraphics[width= 0.45\textwidth]{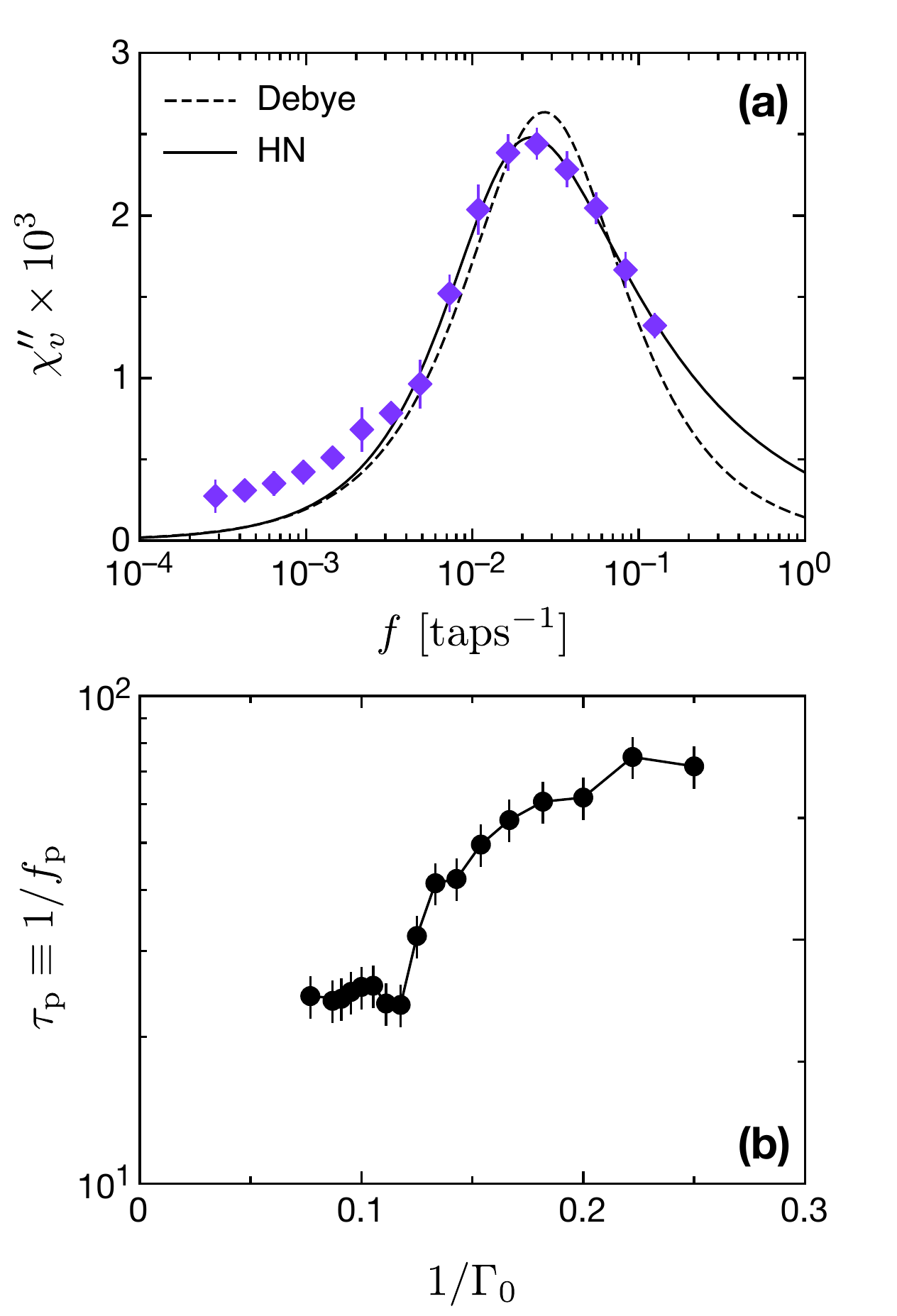}
\caption[Extracting characteristic relaxations times from $\tilde\chi_{v}(f)$]{(Color online) (a) The spectral peak in $\chi_{v}''$ at $\Gamma_{0}=7$. Lines are fits to the spectral peak using Debye (broken line) and Havriliak-Negami (solid line) forms. (b) The characteristic relaxation time $\tau_\mathrm{p}=1/f_\mathrm{p}$ plotted against $1/\Gamma_{0}$.}
\label{SpectralPeak}
\end{figure}

Fig.~\ref{Spectra} plots the dynamical spectrum $\tilde{\chi}_v(f)$ obtained at different values of $\Gamma_0$. Here, the imaginary part of the spectrum $\chi_v''(f)$ measures the spectral distribution of relaxation processes in granular compaction. In overall form, the spectrum of $\tilde{\chi}_v$ is similar to its dielectric and magnetic counterparts in glasses, including a peaked imaginary part. As with glasses, the spectral peak in $\chi_{v}''$ is wider than that of a single Debye peak. In the vicinity of the peak frequency, $\chi_{v}''$ it can be fitted to the generalized Havriliak-Negami form \cite{HavriliakNegami1966}
\begin{equation}
\tilde{\chi}_v(f) = \tilde{\chi}_v^\infty + \frac{\tilde{\chi}_v^0 - \tilde{\chi}_v^\infty}{[1+ (i 2\pi f \tau)^\alpha]^\beta},
\label{HNform}
\end{equation} 
with $\alpha \approx 1$ and $\beta\approx 0.5$ (note $\alpha=\beta=1$ recovers the Debye form) [Fig.~\ref{SpectralPeak}(a)]; however, the low frequency wing of $\chi_{v}''$ is in excess of what a generalized, single-peak form (such as Eq.~\ref{HNform}) can account. For very small $\Gamma_{0}$, the response of $v(t)$ to low-frequency amplitude-modulated tapping can be strongly anharmonic. This can result in anomalous features in the $\tilde\chi_{v}$ spectrum for small $\Gamma_{0}$ at low $f$, such as negative values of $\chi_{v}''$.

It is clear that while the spectral peak in $\chi_v''(f)$ varies in height and shape with $\Gamma_0$, the peak location is nearly stationary. This is in sharp contrast to the dielectric and magnetic spectra of glasses, where a small reduction in $T$ can shift the spectral features to dramatically lower frequencies \cite{Menon&c1992, Reich&c1990, Quilliam&c2008}. By fitting the spectral peak in $\chi_v''$ to Eq.~\ref{HNform}, the peak frequency $f_\mathrm{p}$ (and equivalently, a characteristic relaxation time of the system $\tau_\mathrm{p} \equiv 1/f_\mathrm{p}$) can be extracted. Fig.~\ref{SpectralPeak}(b) plots the inverse peak frequency $1/f_\mathrm{p} \equiv \tau_\mathrm{p}$ versus $1/\Gamma_0$ obtained over a broad range of $\Gamma_{0}$. For $\Gamma_{0}>7.5$, $\tau_{\mathrm{p}} \approx 25$ taps does not change with $\Gamma_{0}$; below $\Gamma_{0}=7.5$, $\tau_\mathrm{p}$ grows modestly from $\tau_\mathrm{p} \approx 25$ to 75 taps, and appears to saturate for very low $\Gamma_{0}$. There is no indication that $\tau_\mathrm{p}$ diverges as $\Gamma_0\rightarrow 0$. 
 
The characteristic relaxation time of $\tau_{\mathrm{p}}\approx 25-75$ taps can only be described as remarkably short. Even at the largest $\Gamma_{0}$ within this range, an loosely-packed initial state will take thousands of taps to reach a steady state. Below $\Gamma_{0}\approx\Gamma^{*}\approx11.5$, when the steady state is practically inaccessible, the characteristic timescale  $\tau_\mathrm{p}$ extracted from $\chi_{v}''(f)$ is at least 3 orders of magnitude smaller than the time it takes for a loosely-packed initial state reach the steady state. The behavior of $\tau_{\mathrm{p}}$ with $\Gamma_{0}$ in granular compaction is in sharp contrast to the analogous behavior of glasses, inverse peak frequency of the dielectric and magnetic susceptibilities diverge as $\tau_{\mathrm{p}}\sim e^{1/T}$ or even faster. 

\subsection{An anomalous ``susceptibility''}

The behavior of $\tilde\chi_{v}$, insofar as it describes the responses of a system with ``glassy'' dynamics, is highly anomalous. The weakness of its aging behavior, as well as the absence of a diverging timescale in its spectrum, render $\tilde{\chi}_{v}$ sharply distinct from the familiar dielectric and magnetic susceptibilities of structural and spin glasses. Strangely, $\tilde\chi_{v}$ does not appear to capture the glassy dynamics of granular compaction: there is a broad separation of scale between the characteristic timescale, extracted from the dynamical spectrum, and the time it takes loosely-packed  initial state to arrive at steady state.

It may be that the anomalous behavior of $\tilde\chi_{v}$ is a unique feature of granular compaction, and has no counterparts in structural or spin glasses glasses. In thermal systems, a susceptibility is strictly defined by the response of a macroscopic thermodynamic observable with respect to its conjugate control parameter or field, e.g. $\chi_{M} = \partial M/\partial h$. For granular compaction, where there is no well-defined statistical mechanics, no principle dictates $\tilde\chi_{v}$ should be the analog of any thermodynamic susceptibility. But as I will suggest in Sec. III, the distinctive features of  $\tilde\chi_{v}$ may not be so anomalous after all; they may instead represent the generic behavior of the configuration specific heat in glassy systems.

\section{Volumetric susceptibility as a ``specific heat''}

Granular compaction is an athermal system with strongly dissipative dynamics; nevertheless, I suggest tapping plays the role in granular compaction akin to temperature in glasses. Most importantly, tapping activates transitions from one static configuration of particles to another, thereby allowing the system to explore configuration space; this is reminiscent of the temperature-driven exploration of the phase space in thermal systems. Additionally, in granular compaction, the tapping amplitude $\Gamma_{0}$ is the principal control parameter that determines whether the dynamics is glassy, while in glasses the analogous role is taken by $T$.

With this in mind, I propose the following analogy: taking $\Gamma_0$ as a temperature-like control parameter, with $v$ playing the role of energy density as proposed by Edwards \textit{et al.} \cite{EdwardsOakeshott1989, EdwardsGrinev1998}, then $\tilde{\chi}_{v}$ should thought as the complex ``specific heat'' of granular compaction. While this analogy cannot be justified from first principles, operationally it proves to be very helpful towards understanding the experimental results.

\subsection{Connection to specific volume fluctuations}

In thermal equilibrium, the specific heat describes the energy (or equivalently, entropy) fluctuations of the system via the Fluctuation-Dissipation Theorem (FDT). Even though a tapped granular packing is not an equilibrium thermal system, it is reasonable to suppose Onsager's Regression Hypothesis --- that the system does not distinguish between spontaneous fluctuations and those induced by a small external perturbation --- should still hold to some approximation, particularly as the experiments described here are largely in the linear response regime \cite{Sethna2009}. If $\tilde\chi_{v}$ is a specific heat-like quantity, with $v$ playing the role of energy density, then $\tilde\chi_{v}$ should also describe steady-state volume fluctuation in granular compaction. Modeling after the FDT between specific heat and energy fluctuations \cite{NielsenDyre1996, Nieuwenhuizen1998}, I hypothesize the steady state of granular compaction can be described by an effective FDT
\begin{equation}
S(f) = NT_{\mathrm{eff}}^{2} \chi_v''(f)/f,
\label{FDT}
\end{equation}
where $S(f)$ is the power spectrum of specific volume fluctuations, $N$ is the number of particles in the system, and $T_\mathrm{eff} = T_\mathrm{eff}(f, \Gamma_{0})$ is the effective temperature that characterizes the steady-state fluctuations on a timescale $1/f$. In the simplest case, in exact analogy to FDT in thermal equilibrium, $T_\mathrm{eff}$ should be frequency-independent for $f < 1/\tau_0$, where $\tau_0$ is a microscopic relaxation time.

\begin{figure}[tbp]
\center
\includegraphics[width= 0.45\textwidth]{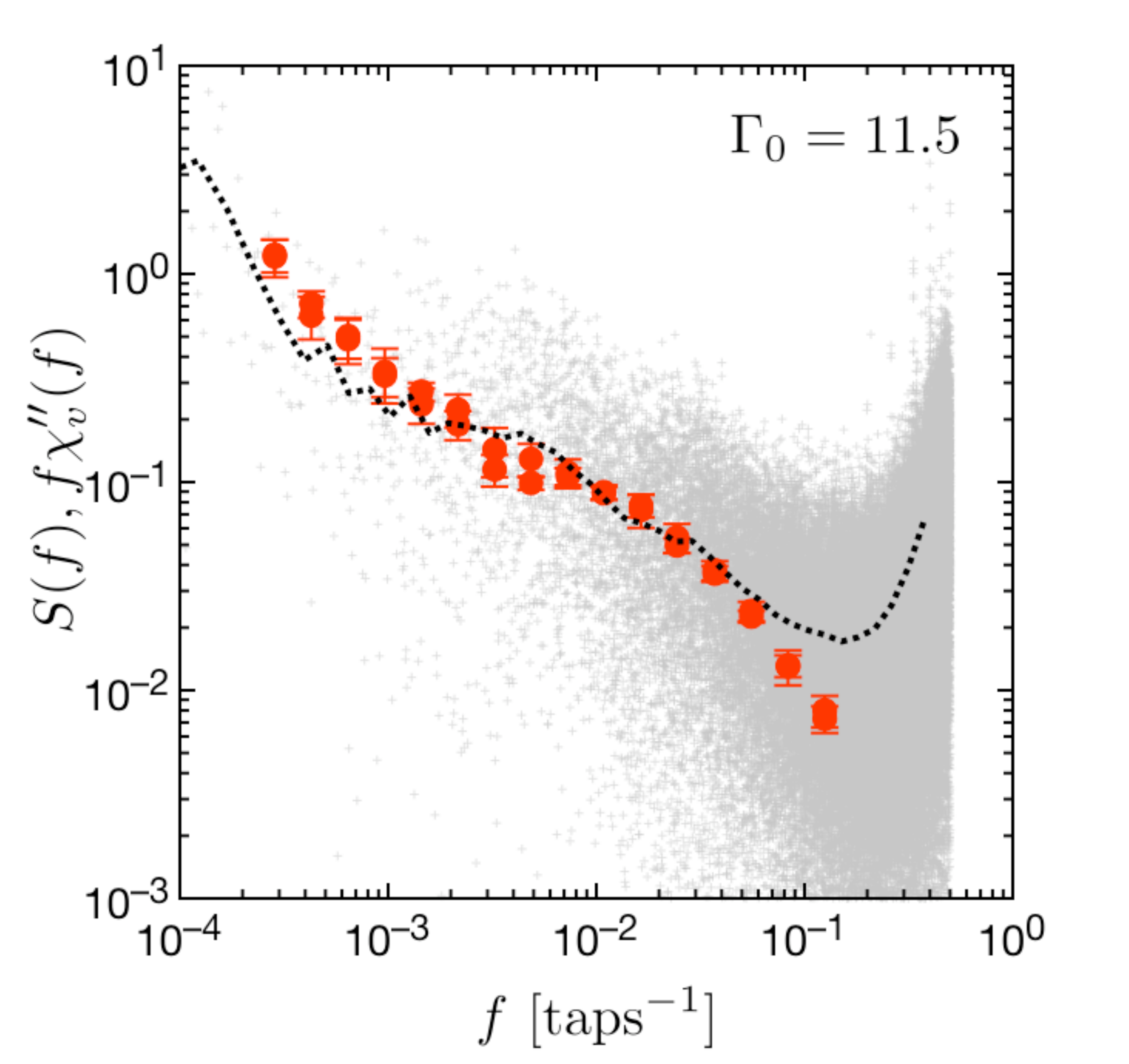}
\caption[FDT-like relationship between $\tilde\chi_{v}$ and volume fluctuations]{(Color online) Testing the effective FDT relationship between volume fluctuations and the volumetric susceptibility at $\Gamma_{0}=11.5$. Plotted are the power spectrum of steady-state volume fluctuations $S(f)$ (light gray crosses), $S(f)$ smoothed by averaging over logarithmically spaced bins (bold broken line), and the combination $\chi_{v}''(f)/f$ (filled symbols).}
\label{FDTplot}
\end{figure} 

The experimental set-up here is not well-suited for measuring volume fluctuations. The bulk-averaging over a large packing of small particles, necessary in order to yield a clean measurement of $\tilde\chi_{v}$, also averages out fluctuations in $v(t)$. In addition, the steady state is only accessible for large $\Gamma_0$, where $\chi_v''$ is small and subject to large uncertainties. The range of $\Gamma_0$ over which all these conflicting demands can be met, and over which Eq.~\ref{FDT} can be tested, is accordingly very narrow. With these caveats in mind, I have independently measured steady-state volume fluctuations and the susceptibility spectrum $\tilde\chi_{v}(f)$ at $\Gamma_{0}=11.5$. Fig.~\ref{FDTplot} plots the power spectrum $S(f)$ of specific volume fluctuations; plotted on top of $S(f)$ is the combination $\chi_v''(f)/f$. While $S(f)$ and $\chi_v''(f)/f$ are clearly distinct at very high frequencies, they are roughly consistent with each other at lower frequencies over 2 decades in bandwidth. This suggests away from the very high frequencies, the relationship between $\tilde\chi_{v}$ and specific volume fluctuations is consistent with that of an effective FDT, with a $T_{\mathrm{eff}}$ that is frequency-independent. The high-frequency discrepancy between $S(f)$ and $\chi_{v}''/f$ could be an artifact caused by tap-to-tap variations in the topography of the top surface, which register as spurious fluctuations in the pack height (and hence specific volume). It may represent an unavoidable departure as $f \rightarrow 1 \ \textrm{taps}^{-1}$, due to the discrete nature of time in granular compaction.

I note here the $S(f)$ shown in Fig.~\ref{FDTplot} has a different slope from that obtained by the earlier study of Nowak \textit{et al.} \cite{Nowak&c1997}. Here $S(f) \sim f^{-1}$, whereas ref.~\cite{Nowak&c1998} measuring at a similar tapping amplitude, found $S(f)\sim f^{-2}$ over the same frequency range \cite{Note1}. The reason for this difference is not clear, but it could due to finite size effects. The system measured in Ref.~\cite{Nowak&c1997} consists of $d=2.2$ mm glass particles confined in a $D=1.8$ cm cylinder ($D/d=8.2$), where the measurement here is of $d=0.5$ mm ceramic particles confined in a $D=2.5$ cm cylinder ($D/d=50$).

\subsection{Comparison with glass-forming liquids}

While the specific is not usually thought as a dynamical quantity, the complex specific heat (at constant pressure) $\tilde{c}_{p} = (c_{p}', c_{p}'')$ has been measured experimentally for a number of glass-forming organic liquids \cite{BirgeNagel1985, Birge1986, DixonNagel1988, Dixon1990}. Approaching $T_\mathrm{g}$ from above, the real part $c_p'(T)$ is nearly flat until close to $T_\mathrm{g}$, when it drops to a lower value over a relatively small span in temperature. Accompanying the stepwise drop in $c_{p}'(T)$, the imaginary part $c_{p}''(T)$ grows from zero far above $T_\mathrm{g}$ to a peak centered at $T_{\mathrm{g}}$, before decaying to zero with further cooling. The spectral evolution of $\tilde{c}_{p}$ with $T$ shows a peak frequency in $c_{p}''(f)$ that shifts rapidly to lower frequencies (indicative of ever larger relaxation times) as $T\rightarrow T_\mathrm{g}$ from above. Qualitatively, the complex specific heat $\tilde{c}_p(f)$ of glass-forming liquids is similar to that of dielectric or magnetic susceptibilities; it does not behave like the volumetric susceptibility $\tilde{\chi}_v(f)$ of granular compaction.

It is possible there exists a low-frequency band in the spectrum of $\tilde{\chi}_{v}$, outside of the experimental window, which does slow down as $\Gamma_{0} \rightarrow 0$. This low-$f$ band could then be analogous to the spectral features observed in the complex specific heat of glass-forming liquids, and would be responsible for the glassy character of granular compaction. In a molecular dynamics study of liquid $\textrm{SiO}_2$ approaching $T_\mathrm{g}$ from above, Scheidler \textit{et al.} \cite{Scheidler&c2001} found the spectrum of the constant-volume specific heat $c_v''(f)$ consists of two bands, a high-$f$ band which does not shift with temperature, and a low-$f$ band which shifts to lower frequencies as $T$ decreases. There, the high-$f$ band comes from the vibrational motion of molecules about a metastable configuration, while the low-$f$ band is contributed by configurational rearrangements. It may be that the spectral features observed in granular compaction is analogous to the high-$f$ band of vibrational processes in liquids. But as granular compaction takes place in an athermal system with zero kinetic energy, it is unclear this analogy is appropriate.

\subsection{Comparison with spin glass models}

\begin{figure}[tbp]
\center
\includegraphics[width= 0.45\textwidth]{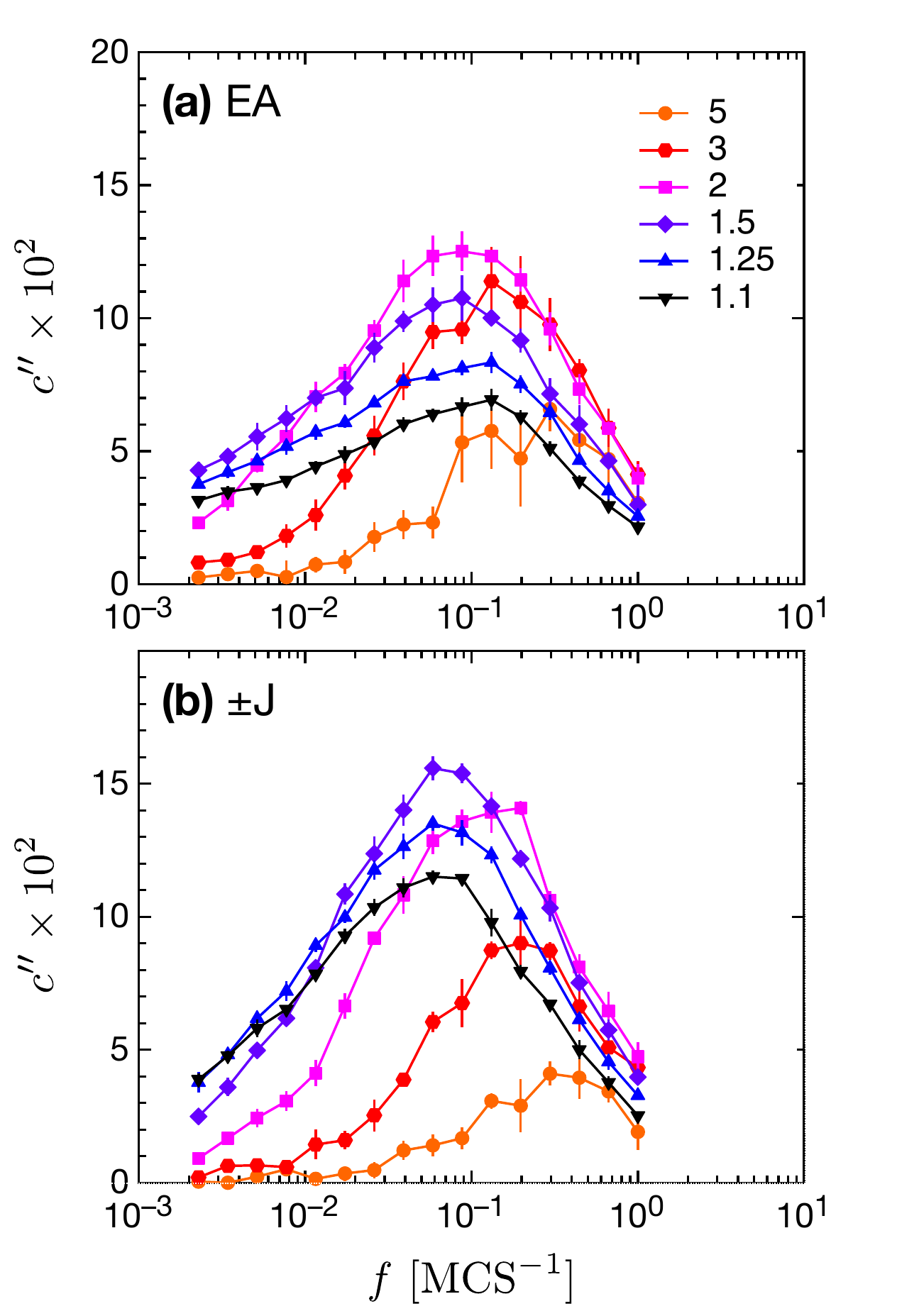}
\caption[The specific heat spectrum of 2D spin glass models]{(Color online) The imaginary part ($c''$) of the spin glass specific heat $\tilde c$ spectrum, obtained at different temperatures $T$ for (a) the Edwards-Anderson (EA) model, (b) the $\pm J$ model. The system size is $16^{2}$. Time is measured in Monte Carlo steps (MCS); 1 MCS = 256 spin flip attempts.}
\label{SGSpectra}
\end{figure} 

There is a crucial distinction between glass-forming liquids and granular compaction in the nature of their entropy. For a glass-forming liquid, its entropy has both configurational contributions (from the spatial arrangement of the molecules) and dynamic contributions (from the kinetic motion of the molecules). The entropy of granular compaction, on the other hand, is purely configurational. Because specific heat is intimately connected to energy and entropy, this difference between glassy-forming liquids and granular compaction may prove crucial. Therefore it may be more appropriate then to compare $\tilde{\chi}_{v}$ with the specific heat of purely configurational systems, such as spin glasses.

The DC magnetic specific heat $c_\mathrm{mag}$ (where the electronic and phonon contributions to the specific heat have been removed) has been measured for several spin glass systems via scanning calorimetry. These have found $c_\mathrm{mag}(T)$ to take broadly peaked forms, qualitatively similar to $\tilde{\chi}_{v}(\Gamma_{0})$, which in some cases exhibits sub-peaks and ``knees'' \cite{MirzaLoram1985, Nagata&c1980, Takano&c2002}. However, being DC measurements, these contain no dynamical information against which one can make a detailed comparison.

In the absence of experimental measurements of the complex specific heat in real spin glasses, I have resorted to Monte Carlo computer simulations to study the purely configurational complex specific heat $\tilde{c}=(c', c'')$ in two standard spin glass models. I examined 2D Ising models governed by the Hamiltonian
\begin{equation}
H = \sum_{(i, j)} J_{ij}S_{i}S_{j} - h\sum_{i}S_{i} \equiv H_{0} -hM,
\end{equation}
where $S_{i} = \pm 1$ are Ising spins occupying the sites of a $16\times 16$ square lattice with periodic boundary conditions. Here $h$ is an externally applied magnetic field, and $J_{ij}$ are quenched random couplings between nearest neighbor spin pairs $(i, j)$. For the Edwards-Anderson (EA) model, $\{J_{ij}\}$ are distributed in two unit-variance gaussians centered at $J = \pm 1$; for the $\pm J$ model, $J_{ij}$ takes on values of  $\{-1, 1\}$ with equal probability.

\begin{figure}[tbp]
\center
\includegraphics[width= 0.45\textwidth]{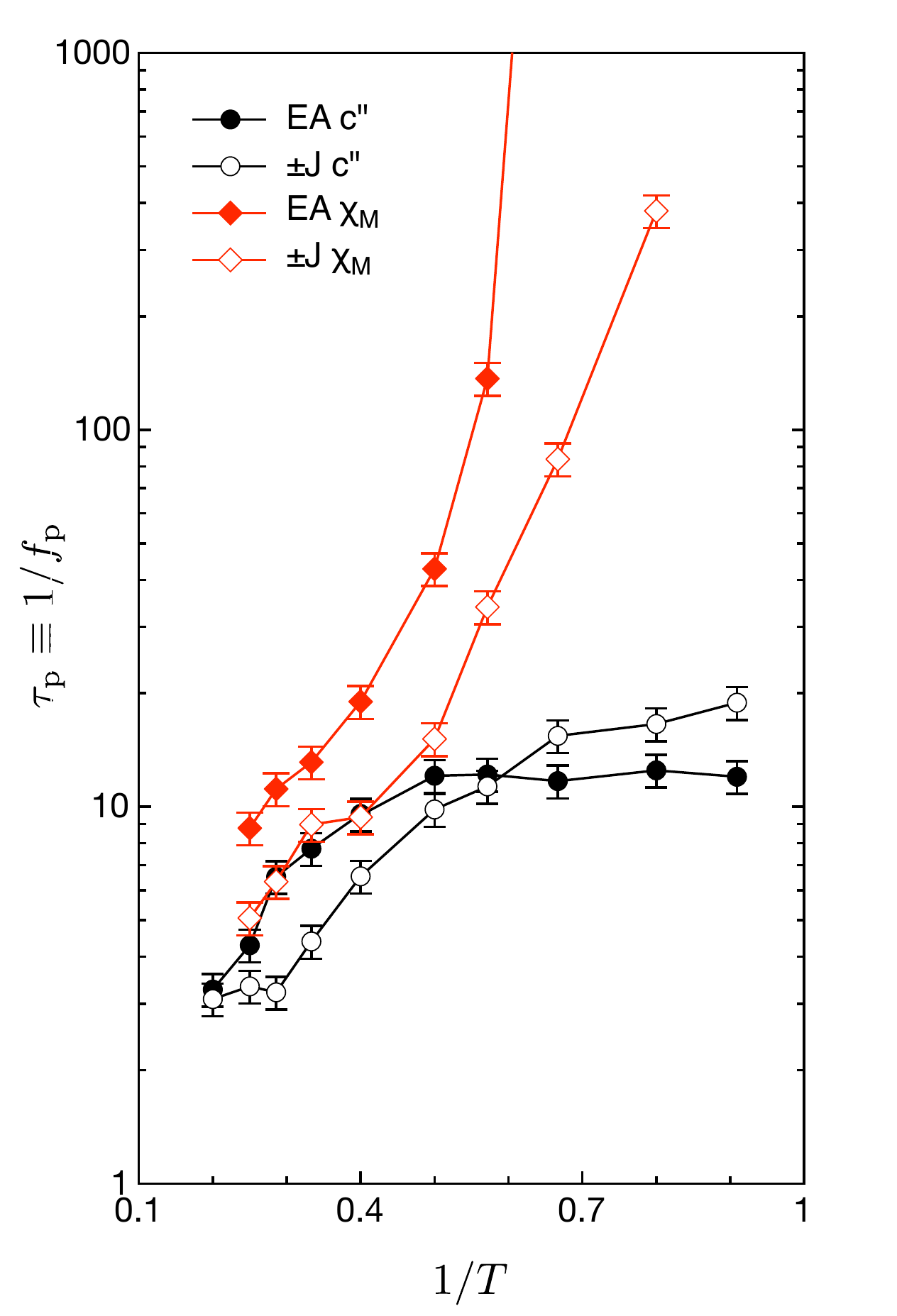}
\caption[Separation of relaxation timescales in spin glass models]{(Color online) The characteristic energy relaxation time $\tau_\mathrm{p} = 1/f_\mathrm{p}$, where $f_\mathrm{p}$ is the peak frequency in $c''(f)$, for the EA and $\pm J$ models (circles). Also shown are magnetic relaxation times obtained from the magnetic susceptibility spectrum for the same models (diamonds). As $T$ is decreased, there is a rapidly growing separation of timescale between energy (obtained from specific heat) and magnetic (from the magnetic susceptibility) relaxation times.}
\label{SGSpectraShift}
\end{figure} 

Fig.~\ref{SGSpectra} plots the spectrum of the imaginary specific heat $c''(f)$ at different temperatures for both the EA and $\pm J$ model. For both models, the height and the shape of the spectral peak varies with $T$, but the inverse peak frequency $1/f_\mathrm{p}\equiv \tau_\mathrm{p}$ grows only modestly and does not appear to diverge as $T\rightarrow 0$. Qualitatively, the overall behavior of $c''(f)$ in both models is similar to that of $\chi_{v}'(f)$, and the forms of $\tau_\mathrm{p}(1/T)$ in the spin glass models are remarkably close to $\tau_\mathrm{p}(1/\Gamma_0)$ as observed in granular compaction (Fig.~\ref{SGSpectraShift}). I have also found at low temperatures, $\tilde{c}$ shows a weaker aging effect than both the overall energy $H$ and the magnetic susceptibility $\tilde{\chi}_M$, reminiscent of the weak aging signature in the volumetric susceptibility $\tilde{\chi}_v$. Thus apparently distinctive features of $\tilde\chi_{v}$ --- weak aging effects and the absence of a diverging timescale --- also appears to characterize the complex specific heat of purely configurational spin glasses.

\subsection{Separation of timescales}

Like the volumetric susceptibility $\tilde\chi_{v}$ in granular compaction, the complex specific heat is characterized by short characteristic timescales, and  does not appear to capture the glassy dynamics of the spin glass models studied. On the other hand, the magnetic susceptibility in the same spin glass models are characterized by rapidly growing timescales that tend to diverge as $T\rightarrow 0$ (Fig.~\ref{SGSpectraShift}). This shows on cooling, there is a rapid separation of timescales between characteristic energy/entropy fluctuations, which remain relatively fast, and magnetic fluctuations, which slow down enormously. However, all dynamical processes in the spin glass, including magnetic fluctuations, must ultimately contribute towards the specific heat; that the spectrum of $\tilde c$ shows no indication of a rapidly growing characteristic timescale suggests the glassy aspects in the low-$T$ dynamics of spin glass must be governed by statistically rare processes.

Is a similar effect at work in granular compaction? As $\tilde\chi_{v}$ appears to describe volume fluctuations via an effective FDT, then the fact that its spectrum shows no sign of a diverging timescale suggests the tapping-induced volumetric compaction of granular packing cannot be understood on the basis of typical relaxations of typical volume fluctuations. If $\tilde\chi_{v}$ can be thought as a specific heat, it should capture all the dynamics of granular compaction. That one sees no indication of a diverging characteristic timescale in $\chi_{v}''$ suggests the glassy aspects of granular compaction are controlled by statistically rare processes which become increasingly separated in timescale from typical relaxation processes as $\Gamma_{0}$ is reduced, leading eventually to the breakdown of ergodicity \cite{Palmer1982, Bouchaud1992}.

\section{The effect of system size}

Specific heat is an intensive quantity that should be independent of system size. But when a system is confined with one or more very small dimensions, its behavior can become qualitatively different from the bulk, and normally intensive quantities can become sensitive to the system size. Most obviously, the properties of a small system with large surface-to-volume ratio will be dominated by that of the interface, which may be qualitatively different from that of the bulk. More subtly, if collective effects are important to the dynamics of the system, then a small system may be unable to support those collective modes characterized by large dynamical correlation lengths, and the contribution of such modes to the specific heat will be lost.

Growing evidence from both simulations and experiments suggest that the collective motions of many particles play a crucial role in the low temperature dynamics of structural glasses. In that case, finite size and confinement effects should modify the glassy state in nontrivial ways. As a model ``glass'' where the small system size is readily accessible, granular compaction is almost ideally suited for looking into how finite size effects can glassy behavior. The volumetric susceptibility $\tilde\chi_{v}$, as a specific-heat-like quantity, is then a potentially sensitive indicator of whether collective particle rearrangements are important to the compaction dynamics.

\subsection{Rescaling the system: the effect on $v$ and $\tilde\chi_v$}

By packing particles of different diameters $d$ into a cylindrical cell whose diameter is fixed at $D=2.5\ \textrm{cm}$, the system is effectively rescaled as $D\rightarrow D/d$.  Here, three different particles sizes were used: $d=0.5$, $1.1$, and $1.8$ mm; this gives us packings with rescaled diameters $D/d = 50$, $22.7$, and $13.9$. Since the height ($\approx 40$ cm in all cases) of the packing is much larger than the cell diameter, $D/d$ is the smallest system dimension and should be principally responsible for any finite size effects.

\begin{figure}[tbp]
\center
\includegraphics[width= 0.45\textwidth]{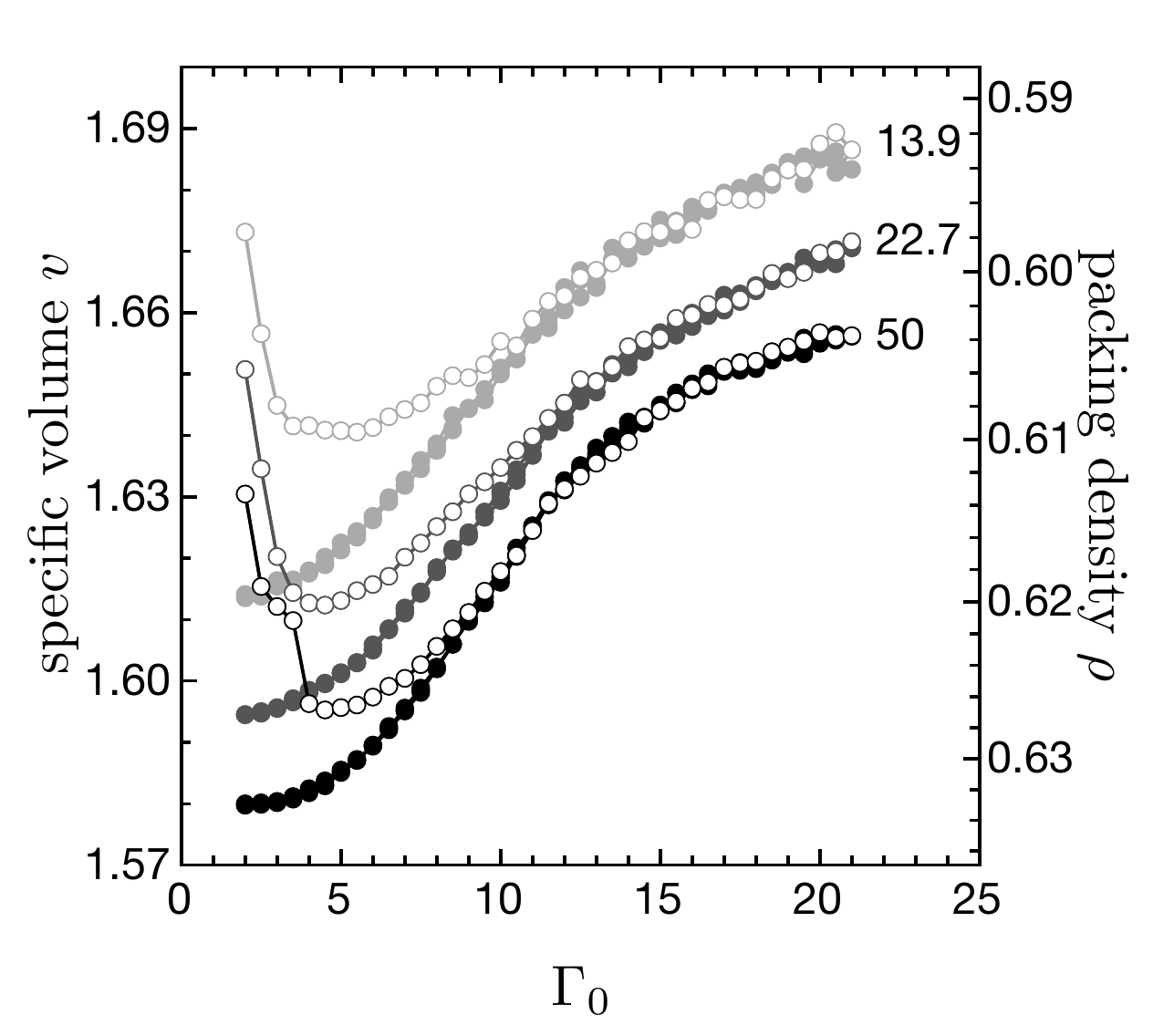}
\caption[The effect of finite system size on $v(\Gamma_{0})$]{The compaction curve $v(\Gamma_{0})$, showing both irreversible (open symbols) and reversible branches (filled symbols), obtained for packings with $D/d=50$, $22.7$, and $13.9$. The ramping of $\Gamma_{0}$ is in steps of $\Delta\Gamma_{0}=0.5$, $\tau_\mathrm{dw} =500\ \textrm{taps}$.}
\label{v(Gamma)}
\end{figure} 

Fig.~\ref{v(Gamma)} plots the volumetric curve $v(\Gamma_{0})$ for all three system sizes $D/d$, obtained using the same ramping, $\Delta\Gamma_{0}/\tau_\mathrm{dw} = 0.5/500\ \textrm{taps}$. Aside from an overall shift to larger $v$ (lower $\rho$), the $v(\Gamma_{0})$ curve for small $D/d$ is not sharply different from that for large $D/d$. The reversible branch is slightly less steep, and the branch point $\Gamma^{*}$ is less clearly defined, but on the whole it is not obvious that any drastic changes have taken place.

\begin{figure*}[tbp]
\center
\includegraphics[width= 0.9\textwidth]{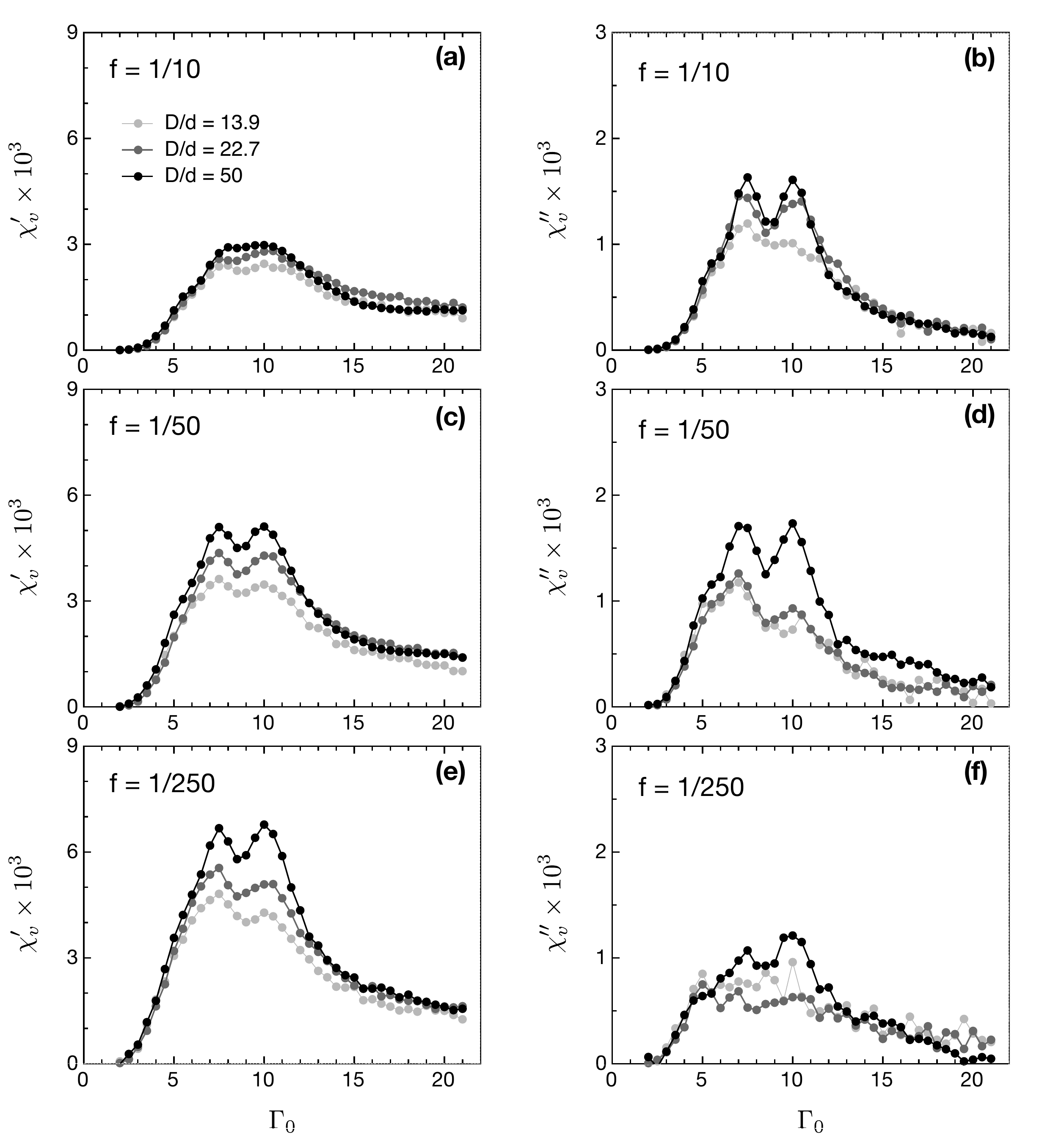}
\caption[The effect of finite system size on $\tilde\chi_{v}(\Gamma_{0})$]{The effect of finite size on the $\Gamma_{0}$-dependence of $\tilde\chi_{v}$ measured at frequencies of   (a, b) $f=10^{-1}$ taps$^{-1}$, (c, d) $f=50^{-1}$ taps$^{-1}$, and (e, f) $f=250^{-1}$ taps$^{-1}$. The systems sizes are $D/d=50$ (black), $D/d=22.7$ (gray), and $D/d=13.9$ (light gray).}
\label{chi(Gamma)FiniteSize}
\end{figure*}

The differences stand out far more clearly when $\tilde\chi_{v}(\Gamma_{0})$ for different system sizes $D/d$ are compared against each other. As shown in Fig.~\ref{chi(Gamma)FiniteSize}, when the tapping is either very strong or very weak, $\tilde\chi_{v}(\Gamma_{0})$ appears to be independent of system size. But in the intermediate regime, where $\tilde\chi_{v}(\Gamma_{0})$ rises to form a broad peak, finite-size effects manifest themselves as a systematic suppression of the overall feature, as well as the sub-peaks, with decreasing $D/d$. Over this intermediate range of $\Gamma_{0}$, a large system is more sensitive, and responds more strongly to small variation in the tapping amplitude, than a small system.

\begin{figure*}[tbp]
\center
\includegraphics[width= 0.9\textwidth]{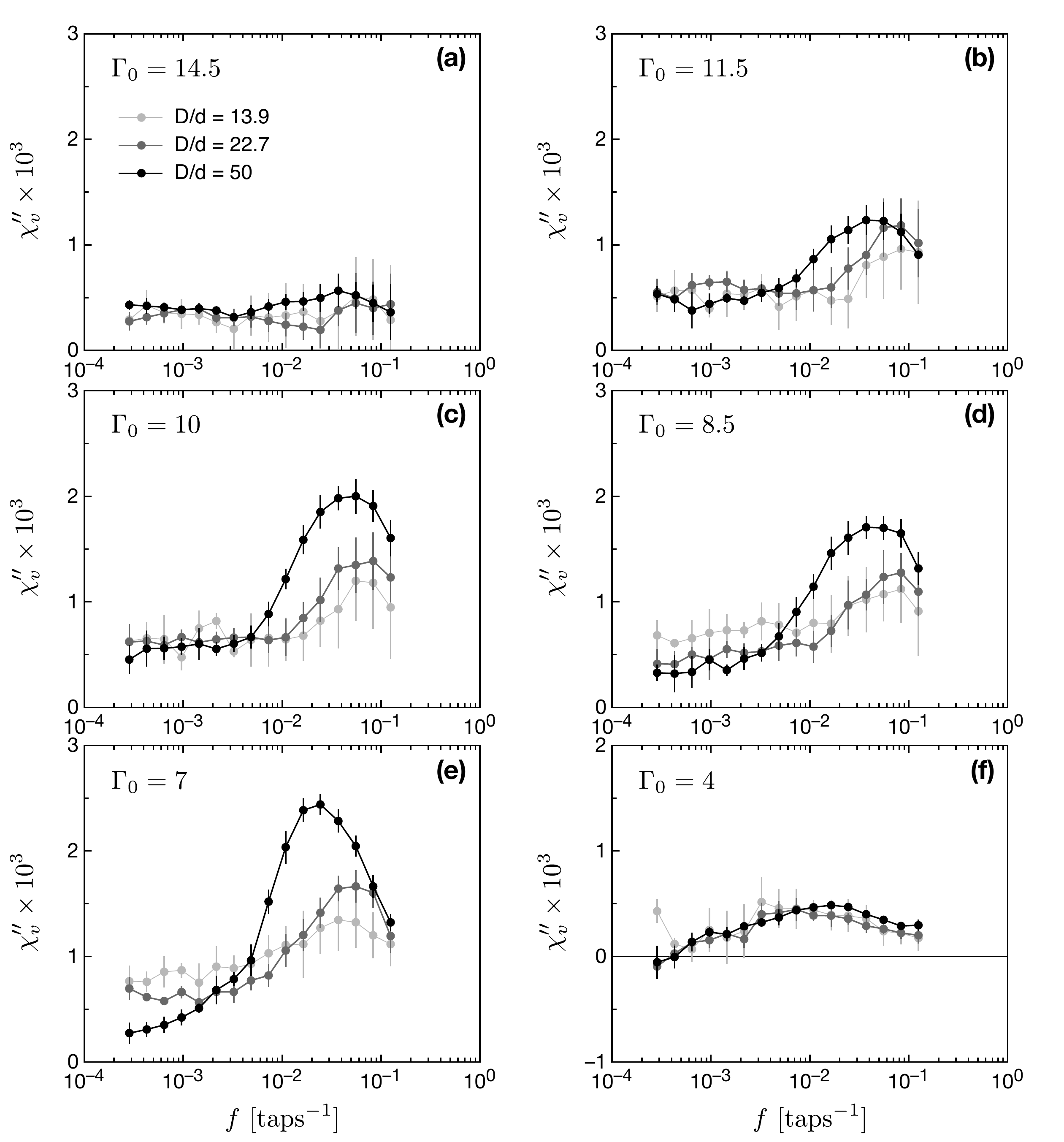}
\caption[The effect of finite system size on the spectrum of $\tilde\chi_{v}$]{The effect of finite size on $\chi_{v}''(f)$, the imaginary part of $\tilde\chi_{v}$ spectrum, measured at (a) $\Gamma_{0}=14.5$, (b) $\Gamma_{0}=11.5$, (c) $\Gamma_{0}=10$, (d) $\Gamma_{0}=8.5$, (e) $\Gamma_{0}=7$, and (f) $\Gamma_{0}=4$. The systems sizes are $D/d=50$ (black), $D/d=22.7$ (gray), and $D/d=13.9$ (light gray).}
\label{chi(f)FiniteSize}
\end{figure*}

Just as finite-size effects appear most prominently within an intermediate range of $\Gamma_{0}$, there is also a window in frequency where it most strongly modifies the spectrum of $\tilde\chi_{v}$. Fig.~\ref{chi(f)FiniteSize} compares the frequency spectra $\tilde\chi_{v}(f)$ at fixed $\Gamma_0$ for different system sizes $D/d$. For large $\Gamma_{0}$, the spectrum of $\tilde\chi_{v}$ is essentially the same for all $D/d$ [Fig.~\ref{chi(f)FiniteSize}(a)]; but as $\Gamma_{0}$ is lowered, the spectral peak in $\chi_{v}''$ becomes increasingly suppressed for smaller systems. The suppression is asymmetric: the low-frequency wing of the spectral peak is more strongly suppressed than the high frequency wing, so that for smaller $D/d$, the peak center is located at higher frequencies [Figs.~\ref{chi(f)FiniteSize}(b, c)]. At still smaller values of $\Gamma_0$, while the spectral peak of a small system remains strongly suppressed, its low-frequency wing decays slower than that of the large system, so that at very low frequencies, $\chi_v''$ for small $D/d$ is systematically enhanced relative to that of large $D/d$ [Figs.~\ref{chi(f)FiniteSize}(d, e)]. Finally, for very weak tapping, the spectrum of $\tilde\chi_v$ once again is essentially the same for all system sizes [Fig.~\ref{chi(f)FiniteSize}(f)]. 

While the finite size effect manifests itself in the specific volume as mostly a uniform shift to larger $v$ (lower density), its effects on $\tilde\chi_{v}$ appears most prominently within windows both in $\Gamma_{0}$ and in $f$. The smaller system has a faster characteristic relaxation time, as denoted by the inverse peak frequency of $\chi_{v}''(f)$, but given the nontrivial changes in the dynamical spectrum with $D/d$, it is difficult to state categorically whether small systems are more or less ``glassy''. What is clear is that the packing with small $D/d$ is not simply a shrunken version of the larger packing --- a qualitative change in the dynamics has taken place.

\subsection{Extrinsic and intrinsic length scales}

For a granular packing confined inside a cylindrical cell, the surface-to-volume ratio scales with the extrinsically imposed cell diameter as $(D/d)^{-1}$. For the smaller system, those processes that take place near the boundary will exert a proportionally stronger influence on its overall behavior. The wall not only restricts the range of motion for particles near the boundary, it also restricts the ways these particles can pack together. This is a particularly strong effect when $D/d$ is small and the curvature of the boundary proportionally large. This geometric hinderance to efficient packing is likely the reason behind the reduced packing density when $D/d$ is small.

At low tapping amplitudes $\Gamma_{0} \leq 7$, it is observed that the particles next to the wall arrange themselves into an ordered triangular layer whose configuration appears to be nearly static from tap to tap. The larger surface-to-volume ratio of small systems means a greater proportion of the particles are semi-immobilized by the wall-induced ordering. The slow rearrangement of particles adjacent to the wall may explain the excess response of small systems at very low frequencies and weak tapping [Figs.~\ref{chi(f)FiniteSize}(d, e)]; but this does not explain the asymmetric suppression of the spectral peak in $\chi_{v}''(f)$.
 
Suppose the tapping-induced particle rearrangements during compaction are collective in nature and can be characterized by a set of correlation lengths $\{\xi\}$ (which will depend on $\Gamma_{0}$). When the system size is small, those responses characterized by $\xi > D/d$ will be suppressed. Since it is not possible to track individual particle motion in these experiments, the presence of spatially correlated particle rearrangements cannot be directly ascertained; one must seek instead some indirect probe of the putative length scales intrinsic to the system. Here I will rely on the following ansatz: at a given $\Gamma_{0}$, it is the collective rearrangement of the particles on a large length scale that contributes to the low-$f$ responses, while small scale rearrangements give rise to the high-$f$ responses. In that case the correlation lengths can be mapped onto frequencies, $\xi\rightarrow\xi(f)$, and the spectral distribution of relaxation process, as measured by $\chi_{v}''(f)$, also describes the distribution of dynamic correlation lengths $\xi$.

The asymmetric suppression of the spectral peak in $\chi_{v}''$, when $D/d$ is small, can then be understood as the suppression of large-scale collective rearrangements. Assuming that for a small system with strong finite size effects, the low-$f$ roll-off in the spectral peak takes place at $f=f_{D}$, corresponding to when $\xi(f_{D}) = D/d$, then a rough bound may be placed on the most probable correlation length $\xi_{\mathrm{p}}$ (corresponding to the peak frequency $f_{\mathrm{p}}$) for collective rearrangements in the largest system. From this admittedly crude bound, I estimate that for $D/d=50$, $\xi_\mathrm{p}$ is not more than $\sim20$ particle diameters. Since finite size effects are only apparent in an intermediate range in $\Gamma_{0}$, $\xi$ must have a non-monotonic variation with $\Gamma_{0}$; $\xi$ must be smaller than the smallest $D/d$ (=13.9) when the tapping is either very strong and very gentle, while reaching some maximum size at intermediate values of $\Gamma_{0}$ that mark the sub-peaks in $\tilde\chi_{v}(\Gamma)$.

\subsection{Comparison with glasses}

While some aspects of the modified volumetric response in small systems can be attributed to the larger surface-to-volume ratio, the most prominent effect --- the asymmetric suppression of the spectral peak in $\chi_{v}''$, suggests a small system lacks some of the responses found in systems with a larger $D/d$, particularly if $\tilde\chi_{v}$ is taken to be the ``specific heat'' of granular compaction. Presumably, these missing responses come from collective particle rearrangements characterized by correlation lengths $\xi > D/d$. From the variation in the finite size effect with tapping amplitude, these correlation lengths should have a non-monotonic $\Gamma_{0}$-dependence. $\xi$ should be small when the tapping is either very strong or very weak, and should grow to maximum size at some intermediate $\Gamma_{0} \approx 10$ where the volumetric response is strongest.

These conclusions agree qualitatively with what we know about dynamic correlations in supercooled liquids. Molecular dynamics simulations of supercooled liquids have found string-like clusters of mobile particles that rearrange collectively; the characteristic sizes of these clusters grows from $\sim 5$ particle diameters far above $T_\mathrm{g}$  to $\sim 20$ particle diameters just above $T_\mathrm{g}$ \cite{Donati&c1998, Donati&c1999}. Experimental studies of glass-forming liquids using confinement within a nano-porous substrate \cite{Arndt&c1997, Alba-Simionesco&c2003}, specific heat spectroscopy \cite{DonthHuthBeiner2001}, multi-point susceptibilities \cite{Bethier&c2005}, and NMR \cite{Reinsberg&c2001} all point indirectly towards growing dynamic correlations on the scale of tens of molecular diameters as $T \rightarrow T_\mathrm{g}$. Collective rearrangements have been directly observed in colloidal glasses where the glass transition is controlled by the colloidal volume fraction $\phi$. Approaching the glass transition point $\phi_{\mathrm{g}}$, the correlated clusters grow in size from a few particle diameters to a few tens of particle diameters, but the decrease rapidly as the system enters deeply into the glassy state ($\phi > \phi_{\mathrm{g}}$)\cite{Weeks&c2000, CuiLinRice2001}. In granular systems, string-like clusters of dynamically correlated particles, on scale $\sim10$ particle diameters, have been directly observed in granular beds fluidized by gas flow; there the clusters grow in size as the external driving is turned down \cite{Keys&c2007}. Self-organized clusters have also been observed in 2D systems composed of particles that slowly swell in size and jam, but are otherwise not driven by external forces; here the clusters were found to grow in size as the system approaches the jammed state \cite{Cheng2009}.

\section{Discussion and outlook}

In this paper, I have described the experimental study of tapping-induced granular compaction, using a volumetric spectroscopy to directly probe the distribution of relaxation processes. This extracts a complex ``susceptibility'' $\tilde\chi_{v}$, which describes how the specific packing volume $v=1/\rho$ responds to small variations in the tapping amplitude $\Gamma$ about a mean level. The principal goal of this work is to make a detailed comparison between the dynamics of granular compaction with the slow relaxations of structural and spin glasses.

\subsection{Glassy state in compaction dynamics} 

Despite some qualitative similarities, the volumetric susceptibility $\tilde\chi_{v}$ is sharply distinct from its dielectric and magnetic counterparts in structural and spin glasses in important ways. $\tilde\chi_{v}$ exhibits very weak signs of aging, and its frequency spectrum shows no indication of a rapidly slowing relaxation time. There is a large separation between te characteristic relaxation time extracted from $\tilde\chi_{v}$, and the time it takes a loose-packed initial state to reach steady state. The slow, glassy dynamics of granular compaction simply does not appear to manifest itself in the spectral response.

Monte Carlo simulations of two standard spin glass models show that the complex specific heat $\tilde c$ in these models bear striking similarities to $\tilde\chi_{v}$. In particular, as $T$ is reduced, the spectrum $\tilde c$ shows only a modestly growing relaxation time in a manner that is remarkably similar to the behavior of $\tilde\chi_{v}$ when $\Gamma$ is turned down. This suggests the features of $\tilde\chi_{v}$ are not distinctive to granular compaction, but may be generic to the specific heat in purely configurational glassy systems. This idea if further supported by the relationship between $\tilde\chi_{v}$ and volume fluctuations, which appears to obey an effective FDT in analogy to the relationship between specific heat and energy fluctuations in thermal systems. Interestingly, granular compaction appears to split the difference between structural and spin glasses: structurally it is similar to former, while dynamically it behaves like the latter.

If $\tilde\chi_{v}$ can be taken as a specific heat, then it should contain contributions from all relaxation processes, including those responsible for glassy behavior. That a diverging timescale cannot be discerned from the spectrum of $\tilde\chi_{v}$ (or from the specific heat in spin glass models) suggests glassy behavior is largely controlled by a small minority of relaxation processes. As $\Gamma_{0}$ or $T$ is reduced, these rare glassy processes become rapidly separated in timescale from typical relaxations of the system, leading eventually to broken ergodicity and the departure from equilibrium. 

If my interpretation here of finite size effects in $\tilde\chi_{v}$ are correct, it suggests that the glassy dynamics of compaction is governed by rare events and characterized by a distribution of dynamical correlation lengths. These appear to point toward a ``droplet'' picture of the glassy state in granular compaction. For structural glasses, the droplet model suggests the glassy state is a mosaic structure composed of small domains (droplets) drawn from a family of disordered equilibrium thermodynamic phases; the surface energy of the droplets scales weakly with droplet size, so that the droplets grow slowly and no single droplet can dominate the system \cite{LubchenkoWolynes2004, LubchenkoWolynes2007}. Most of the relaxation processes in the system describe equilibrium droplet fluctuations, and have a correlation length on the scale of droplet size. However, droplets also grow or shrink via rare, activated events, and these are the processes responsible for glassy dynamics. The results here suggest a similar droplet picture may be constructed for granular compaction as well.

\subsection{Connection to jamming}

A principal motivation of the present work is to explore the connection between structural glasses and granular packing, as suggested by the Jamming Phase Diagram. Jamming is the consolidation of loose particles into a rigid packing when the density is increased at $T=0$. For frictionless hard spheres, there is a well-defined jamming point $J$ (where $\rho=\rho_{\mathrm{rcp}}$) that exhibits special scaling properties that mark it as an unconventional critical point \cite{OHernSilbertLiuNagel2003, SilbertLiuNagel2005, SilbertLiuNagel2006}. Being a zero temperature phenomenon, the jamming transition does not inherently contain any dynamics. But evidence from experiments and simulations suggest that the critical properties of point $J$ extend beyond its immediate vicinity to influence dynamics at finite-$T$, where point $J$ is not directly accessible, and be felt in a modified form as the glass transition \cite{Zhang&c2009}.

In granular compaction, the ideal transition between jammed and unjammed states at point $J$ is never explored, since any packing one produces experimentally is automatically jammed. Because real particles have friction, instead of a single jamming point at RCP, there is a large configuration space of jammed states, with densities ranging from random loose packing (RLP) at $\rho_\mathrm{rlp}\approx 0.55$ up to $\rho_\mathrm{rcp}\approx 0.64$. Tapping provides the dynamics for the system to explore this configuration space, with point $J$ being an idealized limiting point.

If the influence of point $J$ can extend to finite-$T$ and manifest itself as the glass transition, then a similar influence may also be felt at finite friction and control the transition between glassy and steady state dynamics in granular compaction. I suggest the fuzzy transition between liquid and glass, and the fuzzy transition between steady-state and glassy dynamics in granular compaction, may be thought of as two aspects of the same idealized jamming transition, manifested non-ideally at finite $T$ and finite friction, respectively. This may be why similar dynamical manifestations, such as collective rearrangements on the scales of few tens of particles, are found in both glasses and externally-driven granular systems. In jamming, there appears to be a diverging dynamical correlation length associated with vibrational modes of the system \cite{SilbertLiuNagel2005, WyartNagelWitten2005, WyartSilbertNagelWitten2005}. This diverging length scale may not survive way from zero temperature and zero friction, but its vestiges could be the source of the small-to-modest dynamic correlation lengths in glasses and in granular compaction.

\subsection{Outstanding questions}

What are the rare processes and slow fluctuations that give us the glassy behavior of granular compaction? In spin glasses, it is the magnetic fluctuations that slow down on cooling. In structural glasses, viscosity diverges on cooling, leading to increasingly slow relaxation times for internal stresses. A good candidate for an analog in granular packing are fluctuations in contact force network between particle \cite{CorwinJaegerNagel2005, CorwinHokeJaegerNagel2008}. For very weak tapping, one expects a contact network, once formed, will be minimally perturbed from tap to tap, and that the fluctuations in the contact forces will be characterized by very long relaxation times. In this case, a susceptibility may not be available, since the obvious coupling field, gravity, is not normally controllable; but it should be possible to measure the power spectrum of contact force fluctuations.

What are is role of kinetic degrees of freedom in the glass transition of liquids? Why does the complex specific heat $\tilde c_{p}$ of glass-forming liquids show indications of a diverging timescale, while specific heat of spin glass simulations and the volumetric susceptibility $\tilde\chi_{v}$of granular compaction do not? One way this question may be attacked is to extend volumetric spectroscopy to continuous tapped or shaken systems, where kinetic energy is important. Mechanical spectroscopy using a torsional oscillator on granular packings fluidized by continuous vibrations have found a growing relaxation time as the vibrational amplitude is turned down \cite{DAnnaGremaud2001a, DAnnaGremaud2001b, DAnna&c2003}. It will be interesting to compare the rotational and volumetric susceptibility, which one can take provisionally to be the analogs of viscosity and specific heat, in the same system. Do the timescales for the two processes track each other as the vibrations are turned down, or do they decouple instead? 

What do particle rearrangements in granular compaction look like? In particular, is the ansatz that low-frequency responses are contributed by large-scale particle rearrangements correct? Visualizing particle motions in a 3D packing over the course of an extended experiment will be very challenging. One possibility is to do so on a coarse-grained way using clear particles mixed with a small fraction of opaque tracers. By imaging the pack using backlighting from multiple angles, a rough optical ``tomography'' may be constructed, and collective particle rearrangements may be inferred from frequency-locked cross-correlations in the image field. Larger system with larger $D/d$ should be studied to see when do finite size effects disappear completely. This will give us an upper critical length scale for collective rearrangement in granular compaction. It will be especially interesting if this upper critical length turns out to be very large. Since granular compaction is arguably closer to the ideal $T=0$ jamming transition than either colloidal glasses or glass-forming liquids, it is possible that a diverging length scale associated with point $J$ will leave a larger vestige in granular compaction.

Throughout this paper I have made purely operational analogies between granular compaction and the dynamics of structural and spin glasses. But there is no first-principle justification that the athermal, dissipative packing of grains can be described by statistical mechanics using macrosocopic observables. That the relationship between the volumetric susceptibility and steady-state volume fluctuations appear to obey an effective FDT is consistent with Edward's proposal for granular statistical mechanics \cite{Note2}. The validity of the statistical mechanics approach can be tested in detail if ``susceptibilities'' in addition to $\chi_{v}$ can be proposed and measured. Will these also obey an effective FDT with respect to the relevant fluctuations? If so, is the same effective temperature felt for the different susceptibilities? And how does the effective temperature depend on the strength  and the manner in which the packing is driven? It is useful here to recall that for all but the simplest thermal systems, the basic assumptions of standard statistical mechanics cannot be justified except by how well the theory describes experimental results. Apart from what they can tell us about glassy dynamics, one value in experiments such as the one here is in providing new and varied data that can test, experimentally, the validity of the statistical mechanics approach in describing granular materials. 

\acknowledgements

I am greatly indebted to Sidney Nagel for his support and advice. Through the course of this work, I have benefited from discussions with Tom Witten, Dan Silevitch, Sue Coppersmith, Bob Leheny, Heinrich Jaeger, Simon Swordy, and Scott Wakely. A special acknowledgement is due to Xiang Cheng not only for discussions, but also for his technical assistance with aspects of the experiment. This work is supported by NSF MRSEC DMR-0820054, NSF DMR-0652269 and DOE DE-FG02-03ER46088.

\end{document}